\documentclass{jfm}
\usepackage{graphicx}
\usepackage{epstopdf, epsfig}
\usepackage{amssymb}
\usepackage{amsmath}
\usepackage{gensymb}
\usepackage{xcolor}
\usepackage{mathrsfs}
\usepackage{caption}
\usepackage{subcaption}
\usepackage{mathtools}
\usepackage{longtable}
\usepackage{array}
\usepackage{titlesec}

\titleformat{\paragraph}
{\normalfont\normalsize\itshape}{\theparagraph}{1em}{}
\titlespacing*{\paragraph}
{0pt}{3.25ex plus 1ex minus .2ex}{1.5ex plus .2ex}

\newcolumntype{C}[1]{>{\centering\arraybackslash}m{#1}}

\shorttitle{Tip vortex cavitation mechanism}
\shortauthor{S. Lak and R. Jaiman}

\title{A numerical study on the oscillatory dynamics of tip vortex cavitation}

\author{Saman Lak\aff{1}
 \and Rajeev Jaiman\aff{1}
 \corresp{\email{rjaiman@mech.ubc.ca}}}

\affiliation{\aff{1}Department of Mechanical Engineering, University of British Columbia, Vancouver, BC, Canada}

\begin{document}

\maketitle

\begin{abstract}
In this paper, we numerically study the mechanism of the oscillatory flow dynamics associated with the tip vortex cavitation (TVC) over an elliptical hydrofoil section.
Using our recently developed 3D variational multiphase flow solver, we investigate the TVC phenomenon at a Reynolds number of $Re = 8.95 \times 10^5$ via dynamic subgrid-scale modeling and the homogeneous mixture theory. To begin, we examine the grid resolution requirements and introduce a length scale that considers both the tip vortex strength and the core radius. 
This length scale is then employed to non-dimensionalize the spatial resolution in the tip vortex region and establish the mesh requirements for large eddy simulation of tip vortex cavitation. We next perform simulations to analyze the dynamical modes of tip vortex cavity oscillation at different cavitation numbers and compare them with the semi-analytical solution.
The breathing mode of cavity surface oscillation is extracted from the results through the definition of an effective radius. The temporally-averaged effective radius demonstrates that the cavity experiences a growth region followed by decay as it progresses away from the tip. Further examination of the characteristics of local breathing mode oscillations in the growth and decay regions indicates the alteration of cavity's oscillatory behavior as it travels from the growth region to the decay region, with the oscillations within the growth region being characterized by lower frequencies. 
For representative cavitation numbers $\sigma \in  [1.2,2.6]$, we find that pressure fluctuations exhibit a shift of the spectrum towards lower frequencies as the cavitation number decreases, similar to its influence on breathing mode oscillations.
The results indicate the existence of correlations between the breathing mode oscillations and the pressure fluctuations. It is found that the growth and decay regions contribute differently to the pressure fluctuations. The low-frequency pressure fluctuations are found to be correlated with the growth region, whereas the breathing mode oscillations within the decay region are correlated with higher-frequency pressure fluctuations, where the center frequency of a hump is observed. 


\end{abstract}

\begin{keywords}

\end{keywords}

\section{Introduction}
\label{section:Introduction}
Noise pollution due to human activity (i.e., anthropogenic noise) poses serious threats to the marine ecosystem (\cite{2021_Duarte_Sci}). One of the main sources of anthrophony is shipping, which generates noise through the operation of propellers, hull vibrations, and onboard machinery \citep{2022_Smith_JOE}.
A significant proportion of the propeller noise is due to cavitation phenomenon occurring in various forms such as sheet and tip vortex cavitation. In most cases, tip vortex cavitation is the first form of cavitation appearing on ship propellers (\cite{2015_Zhang_JOH}), which arises due to the low pressure within the vortex trailing from the tip of propeller blades.

Tip vortex forms on lifting surfaces of finite length due to the pressure difference between the pressure side and the suction side which disappears at the tip. The pressure gradient on the pressure and suction sides leads to opposite spanwise velocity components on these surfaces, which generates a vortex trailing from the tip (\cite{1979_Platzer}). The wake sheet shed from the blade rolls up into the tip vortex as flow moves downstream, further strengthening the tip vortex (\cite{1964_Batchelor_JFM}). The pressure within the vortex core may drop below the vapor pressure, leading to a cavitating tip vortex. 
Tip vortex cavitation encompasses a complex dynamical interaction among vortical motion, cavitation, and turbulence (\cite{2002_Arndt_ARFM}).
This complex phenomenon can be observed in a relatively simple hydrofoil section which produces a strong swirling flow with pressure reduction and cavitation process. This paper is motivated by the need to provide an improved understanding of the interplay of cavitation and vortex dynamics.

Understanding the physics of tip vortex cavitation, including the tip vortex roll-up and formation, TVC inception, tip vortex cavity dynamics, and the contribution of tip vortex cavitation to the underwater radiated noise (URN) level in various operating conditions is crucial for the development of effective TVC mitigation strategies. Numerous aspects need to be investigated when the tip vortex flow field is to be studied, even in non-cavitating (wetted) conditions. During the formation of the tip vortex, the tip vortex-boundary layer interactions strongly affect the tip vortex flow field (\cite{1997_Maines_JFE}). Downstream of the tip, the influence of the roll-up of the wake into the tip vortex and the complex system of streamwise vortices shed from the tip needs to be considered (\cite{1996_Devenport_JFM}). Further downstream, viscous decay and the introduction of vortex instabilities play an important role in the breakdown of the tip vortex (\cite{2014_Ganesh_POF}).

The occurrence of cavitation further complicates the phenomenon of tip vortex. These complexities coupled with the high three-dimensionality and anisotropy of the flow with extreme gradients (\cite{2020_Asnaghi_JOE}), the impact of turbulent fluctuations on the cavitation inception (\cite{1992_Arndt_JFE}), and tip vortex wandering which gives rise to the need for special treatments when analyzing the acquired data (\cite{1996_Devenport_JFM}), makes the investigation of tip vortex cavitating flows a complex endeavor. Despite all of these difficulties associated with the investigation of cavitating tip vortex flow, various researchers have studied the tip vortex cavitation phenomenon using theoretical, experimental, and numerical methods focusing on the contributions of TVC to underwater radiated noise.

\begin{figure}
    \centering
    \includegraphics[width=0.6\textwidth]{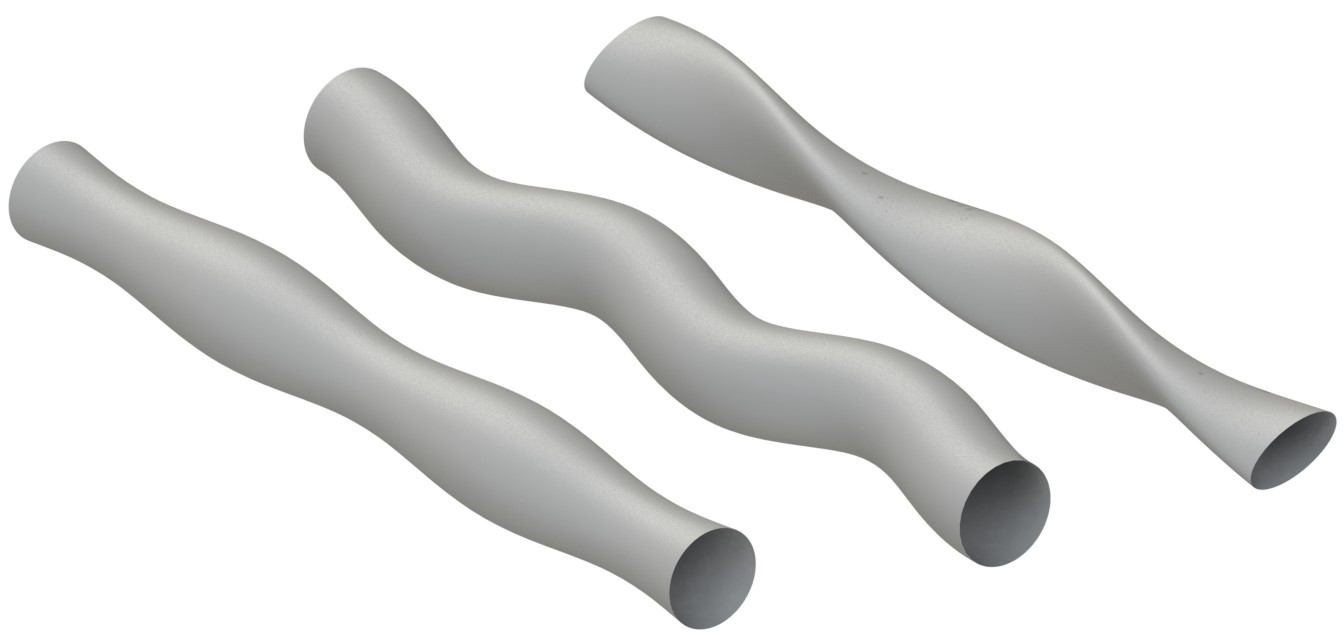}
    \caption{Three main tip vortex cavity oscillation modes: breathing (left), displacement (middle), and double helical (right))}
    \label{fig:CavityModeShapes}
\end{figure}

A significant proportion of the studies aiming to investigate the contributions of TVC to underwater radiated noise place their focus on the oscillations of the tip vortex cavity surface. Different modes of oscillation, namely the breathing mode, serpentine (displacement) mode, and double helical mode, are observed on the tip vortex cavity interface. These cavity oscillation modes are depicted in figure \ref{fig:CavityModeShapes}. These oscillations have been found to affect the radiated noise \citep{Bosschers_thesis_2018, Wang_POF_2023}. An analytical solution for the dispersion relations of the cavity oscillation modes was developed by \cite{Bosschers_thesis_2018}. Experimental evidence for this analytical solution was provided by \cite{2015_Pennings_JFM} by conducting experiments on tip vortex cavitating flow over a stationary NACA66(2)-415 hydrofoil with an elliptical planform. Furthermore, different modes of tip vortex cavitation were observed to induce different levels of instabilities into the tip vortex altering its dynamics \citep{2023_Ye_JFM}. This study investigates the development of cavity surface oscillatory behavior as the tip vortex cavity travels downstream to gain insight into the oscillation dynamics of different regions of the cavity and their contribution to pressure fluctuations in the surroundings.

Due to the limitations of experimental methods in cavitating flow investigations, numerical studies can reveal more details about the flow and cavity characteristics in TVC analysis. Nevertheless, obtaining sufficiently accurate results has been a recent development due to the challenges associated with simulating tip vortex cavitating flows. Large Eddy Simulation (LES) was shown to outperform Reynolds-Averaged Navier-Stokes (RANS) models for simulation of the tip vortex flow field by \cite{Asnaghi_Conf_2017}, and \cite{2020_Asnaghi_JOE} successfully replicated the flow field of tip vortex in wetted conditions observed in the experiments of \cite{Pennings_Experimental} using Implicit LES (ILES). In cavitating conditions, their results regarding the azimuthal velocity distribution and the diameter of the cavitating tip vortex also agreed with the experimental data. They also proposed mesh resolution requirements for accurate simulation of tip vortex cavitating flows in terms of the number of nodes across the vortex core.

To the best of the authors' knowledge, the first numerical study focusing on tip vortex cavity surface oscillation was carried out by \cite{Klapwijk_JOE_2022} using a partially averaged Navier-Stokes and the delayed detached eddy simulation. Their results included some of the cavity surface oscillations, but the breathing mode oscillations were not captured. Moreover, it is evident from the cavity morphology they obtained in their work that the tip vortex cavity was over-dissipated and could not capture a major proportion of the surface oscillations. Recently, the breathing mode cavity surface oscillations were successfully captured in the numerical study by \cite{Wang_POF_2023}. The dominance of the hydroacoustic contribution of the breathing mode was also demonstrated in their work, with the tonal frequency and the center frequency of the broadband hump observed in the noise spectrum being attributed to the breathing mode at zero streamwise wavenumber and the tip vortex cavity resonance. However, their study includes the results for only one cavitation number. Moreover, they only considered the developed tip vortex cavity and not the formation region, i.e., where the wake roll-up into the tip vortex is not complete. In addition, their method for extracting the breathing mode oscillations from the cavity surface oscillation data requires further improvement due to the center of the cavity not being well-defined in tip vortex cavitating flows.

In this study, using LES and the homogeneous mixture model, the tip vortex cavitating flow over a NACA66(2)-415 stationary hydrofoil is numerically investigated via our in-house 3D variational finite element solver. In order to address the adequacy of mesh resolution requirements for the simulation of tip vortex cavitating flows, a length scale is developed based on the radial pressure gradient within the vortex core, which takes into account the strength of the tip vortex as well as its core radius. This length scale is then employed to non-dimensionalize the mesh resolution in the tip vortex region to propose general mesh resolution requirements for the simulation of tip vortex flow, rather than merely considering the number of grid nodes across the vortex core radius. After the assessment of the numerical results with the experimental and analytical data, a comprehensive investigation of the breathing mode of cavity surface oscillations is carried out for three cases with different cavitation numbers. 

To extract the breathing mode oscillations, we define an effective radius based on the cross-sectional area of the tip vortex cavity that captures the cavity volume variations. Different regions the tip vortex cavity experiences as it progresses away from the tip are examined using the temporally-averaged effective radius. The spatially-averaged effective radius is analyzed to discern the temporal variations of the vapor volume of the entire tip vortex cavity.
The spatial-temporal characteristics of local breathing mode oscillations are further investigated to understand the characteristics of this oscillatory mode and its development along the cavity in different cavitation numbers. Proper Orthogonal Decomposition (POD) is employed to further examine the streamwise variations of the breathing mode characteristics. Next, the pressure fluctuations are probed at different locations within the domain in all of the cases to investigate how the contributions of tip vortex cavitation to the pressure fluctuations vary in different cavitation numbers. The spectra of the pressure fluctuations are then compared to the breathing mode oscillation spectra within different regions of the tip vortex cavity to gain insight into the correlations between the local breathing oscillations in different regions of tip vortex cavity and the pressure fluctuations.

The remainder of the paper is structured as follows. In \S \ref{section:ComputationalFramework}, the computational framework employed in this work is described. The problem setup and grid generation strategy are presented in \S \ref{section:ProblemSetup}. The results obtained from the simulations are covered in \S \ref{section:results}. While the focus in \S \ref{section:noncavitating} is placed on the non-cavitating conditions, \S \ref{section:TVCOscillations} provides an overview of the oscillation dynamics of the tip vortex cavity. A systematic investigation of the breathing mode of oscillation is carried out and presented in \S \ref{section:Reff}. \S \ref{section:PressureFluctuations} focuses on the pressure fluctuations in different cavitation numbers and their correlation with breathing mode oscillations. The key findings and concluding remarks are summarized in \S \ref{section:conclusion}.

\section{Computational Framework}
\label{section:ComputationalFramework}
In this work, our in-house high-fidelity finite element flow solver was employed for simulating the tip vortex cavitating flow. The mathematical modeling and numerical formulation are briefly presented here. For further details, the reader is referred to the original papers on the implementation of the models used herein (\cite{Suraj_CAMWA}, \cite{2016_Jaiman_CAF}).

\subsection{Multiphase flow modeling}
In this work, the multiphase flow is treated as a continuous homogeneous mixture of liquid and vapor phases. A phase indicator, $\phi^f(\textbf{x},t)$, determines the phase fraction of the liquid phase at any coordinate and time in the working fluid physical domain, $\Omega^f(\textbf{x},t)$, where $\textbf{x}$ and $t$ are the spatial and temporal coordinates within the domain $\Omega^f(\textbf{x},t)$ with the boundary $\Gamma^f$.
The mixture density ($\rho^f$) and the mixture dynamic viscosity ($\mu^f$) are considered to be linear combinations of the density and viscosity of the liquid and vapor phases and are calculated as

\begin{equation}
    \label{mixdens}
    \rho^f = \rho_l \phi^f + \rho_v (1-\phi^f) \; ,
\end{equation}
\begin{equation}
    \label{mixvisc}
    \mu^f = \mu_l \phi^f + \mu_v (1-\phi^f) \; ,
\end{equation}
where the properties with $l$ and $v$ subscripts represent the properties of the pure liquid and pure vapor phases, respectively.
In order to obtain the value of the phase indicator $\phi^f$ within the domain, a scalar transport equation is solved, which, in its conservative form, is
\begin{equation}
    \label{orderparTE}
    \frac{\partial \phi^f}{\partial t} + \phi^f \nabla \cdot \textbf{u} + \textbf{u} \cdot \nabla \phi^f = \frac{\dot{m}}{\rho_l}  \; \indent \textrm{on} \; (\textbf{x},t) \in \Omega^f \; ,
\end{equation}
where $\textbf{u}$ is the fluid velocity at each coordinate $(\textbf{x}, t)$, and $\dot{m}$ is the finite mass transfer rate from the vapor phase to the liquid phase occurring due to cavitation.

\subsection{Cavitation modeling}
In equation \ref{orderparTE}, the source term representing the mass transfer rate from the vapor phase to the liquid phase $\dot{m}$, which includes both the effect of condensation and evaporation, is calculated using the model proposed by \cite{SchnerrSauer}
\begin{equation}
\begin{aligned}
    \label{SchnerrSauer}
    \dot{m}  & = \frac{3 \rho_l \rho_v}{\rho^f R_B} \sqrt{\frac{2}{3 \rho_l |p^f - p_v|}} \bigg[ C_c \phi^f (1-\phi^f) \max(p^f - p_v, 0) \\
    &\quad + C_v \phi^f (1+\phi_{Nuc} - \phi^f) \min(p^f - p_v, 0) \bigg] \; ,
\end{aligned}
\end{equation}
where $p^f (\textbf{x}, t)$ is the working fluid pressure, and $p_v$ is the saturation pressure of the working fluid. $C_c$ and $C_v$ are the condensation and evaporation coefficients, which were added to the model in later numerical implementations (\cite{Ghahramani_IJMF_2019, Cazzoli_EP_2016}) in order to modify the condensation and evaporation behavior with respect to specific configurations.

In this model, the initial nuclei within the flow are assumed to be of equal radius and homogeneously distributed within the domain.  $\phi_{Nuc}$ in equation \ref{SchnerrSauer} is the initial phase fraction of the bubble nuclei, which is related to the initial volume of the nuclei per unit volume as
\begin{equation}
    \label{PhiNuc}
    \phi_{Nuc} = \frac{v_{nuc}}{1 + v_{nuc}} \; ,
\end{equation}
where $v_{nuc}$ is the initial volume of the nuclei per unit volume, which is related to the nuclei diameter ($d_{nuc}$) and the number of nuclei per unit volume ($n_0$) as $v_{nuc} = \frac{\pi n_0 d_{nuc}^3}{6}$. $R_B$ used in equation \ref{SchnerrSauer} is the equivalent radius of the vapor volume assumed to be in the form of one bubble and is related to other parameters of the flow using the equation
\begin{equation}
    \label{RB}
    R_B = \left( \frac{3}{4 \pi n_0} \frac{1 + \phi_{Nuc} - \phi^f}{\phi ^ f} \right) ^ {1/3} .
\end{equation}

\subsection{Fluid conservation of mass and momentum}
\label{fluidEq}
The spatially filtered Navier-Stokes equations for an incompressible flow are 

\begin{equation}
    \label{NS}
    \rho^f \frac{\partial \overline{\mathbf{u}}} {\partial t} + \rho^f \overline{\mathbf{u}} \cdot \nabla \overline{\mathbf{u}} = \nabla \cdot \overline{\mathbf{\sigma}}^f + \nabla \cdot \mathbf{\sigma}^{sgs} +\mathbf{b}^f \; \; \; \textrm{on}  \; (\mathbf{x},t) \in \Omega^f \; ,
\end{equation}

\begin{equation}
    \label{cont}
    \nabla \cdot \overline{\mathbf{u}} = 0 \; \; \; \textrm{on}  \; \Omega^f \; ,
\end{equation}
where the overbar denotes the spatially filtered quantity, $\mathbf{b}^f$ denotes the body force exerted on the fluid, and the Cauchy and subgrid stresses are represented by $\overline{\mathbf{\sigma}}^f$ and $\mathbf{\sigma}^{sgs}$, respectively.
The Cauchy stress for a Newtonian fluid can be written as
\begin{equation}
    \label{Cauchy}
    \overline{\sigma}^f = -\overline{p} \mathbf{I} + \mu^f \left( \nabla \overline{\mathbf{u}} + \left( \nabla \overline{\mathbf{u}} \right) ^ T \right) \; ,
\end{equation}
where $\overline{p}$ represents the filtered fluid pressure.
The subgrid stress $\sigma^{sgs}$ is the additional stress term appearing in equation \ref{NS} due to the filtering applied in large eddy simulation. In order to complete the set of equations, a model is required for the subgrid stress tensor.

\subsection{Subgrid-scale modeling}
Dynamic subgrid-scale (SGS) model is utilized in this work for modeling the subgrid-scale stress $\mathbf{\sigma}^{sgs}$. In this model, a filter, denoted by an overbar, is applied to the Navier-Stokes equations (\ref{NS}, \ref{cont}), which, as explained in \S \ref{fluidEq}, gives rise to a new stress term, the subgrid-scale (SGS) stress $\mathbf{\sigma} ^ {sgs} _ {ij} = \overline{u_i^f u_j^f} - \overline{u}_i^f \overline{u}_j^f$, to appear in the momentum equation. This stress term corresponds to the stresses originating from the scales of motion smaller than the spatial resolution $\Delta$. As mentioned in \S \ref{fluidEq}, since the SGS quantities are unknown, a model for the subgrid-scale stress is required for the closure of the problem.

The nonlinear SGS stress (\cite{Gatski_JFM_1993, Wang_POF_2005}) is obtained using the equation
\begin{align}
    \begin{split}
        \label{SGS}
        \sigma_{ij}^{sgs} - \frac{\delta_ij}{3} \sigma_{kk}^{sgs} \approx & -2 \mu_t \overline{S}_{ij} - C_{NL} 6 \mu_t^2 / \sigma_{kk}^{sgs} \\
        & \times \left( \overline{S}_{ik} \overline{\Omega}_{kj} + \overline{S}_{jk} \overline{\Omega}_{ki} - 2 \overline{S}_{ik} \overline{S}_{kj} + \frac{2}{3} \overline{S}_{nk} \overline{S}_{kn} \delta _ {ij} \right) \; ,
    \end{split}
\end{align}
where $\mu_t$ is the dynamic eddy viscosity, and $\overline{S}_{ij} \equiv \frac{1}{2} (\frac{\partial \overline{u}_i^f}{\partial x_j} + \frac{\partial \overline{u}_j^f}{\partial x_i})$ and $\overline{\Omega}_{ij} \equiv \frac{1}{2} (\frac{\partial \overline{u}_i^f}{\partial x_j} - \frac{\partial \overline{u}_j^f}{\partial x_i})$ are the resolved strain rate and rate-of-rotation, respectively. The eddy viscosity $\mu_t$ in this equation is related to the mesh resolution and the resolved strain tensor as
\begin{equation}
    \label{eddyVisc}
    \mu_t = \rho^f (C_s \overline{\Delta})^2 |\overline{S}| \; ,
\end{equation}
where $|\overline{S}|$ is the norm of the resolved strain rate tensor calculated as $|\overline{S}| = (2 \overline{S}_{ij} \overline{S}_{ij})^{1/2}$. In order for LES to yield reasonable results close to the wall, a simple algebraic eddy viscosity model (\cite{WallLES1, WallLES2}) is employed instead of equation \ref{eddyVisc} close to the wall as $\frac{\mu_t}{\mu^f} = \kappa y^+_w (1 - e ^ {-y_w^+} / A)$, where $y^+_w = y_w u_\tau / \nu$ is the non-dimensional distance to the wall in wall units with the friction velocity and the model coefficient being denoted by $u_\tau$ and $\kappa$, respectively, and $A$ having a constant value of 19.

In the dynamic SGS model (\cite{MoinSGS}), a test filter is defined in addition to the grid filter. The grid filter, denoted by an overbar, has a scale dimension of $\Delta$, and the test filter, denoted by the operation $\widehat{(\; \;)}$, is any coarser level filter with a scale dimension of $\widehat{\Delta}$ (\cite{Sagaut_2005_LES}). The SGS stress at the test level $T_{ij} = \widehat{\overline{u_i u_j}} - \widehat{\overline{u}}_i  \widehat{\overline{u}}_j$ is obtained by test-filtering the equations of motion. The Leonard stress tensor $\mathscr{L}_{ij}$, which is the stress associated with length scales between the test filter size and the grid filter size
\begin{equation}
    \label{LeonardStress2}
    \mathscr{L}_{ij} = \widehat{\overline{u}_i \overline{u}_j} - \widehat{\overline{u}}_i  \widehat{\overline{u}}_j 
\end{equation}
can be related to $T_{ij}$ and $\sigma^{sgs}_{ij}$ as
\begin{equation}
    \label{LeonardStress}
    \mathscr{L}_{ij} = T_{ij} - \sigma^{sgs}_{ij} \; .
\end{equation}
Using Smagorinsky eddy-viscosity model for the unknows stresses $\sigma^{sgs}_{ij}$ and $T_{ij}$ yields
\begin{equation}
    \label{Lilly}
    \mathscr{L}_{ij} = - 2 C_s^2 \bigg( \widehat{\Delta}^2 \widehat{| \overline{S} |} \widehat{ \overline{S}_{ij} } - \Delta^2 | \overline{S} | \overline{S}_{ij}   \bigg) \; ,
\end{equation}
and the Smagorinsky constant is related to a dynamic coefficient $C_{ds}$ as $C_s = \sqrt{C_{ds}}$ for $C_{ds}>0$, and $C_s=0$ otherwise. The dynamic coefficient $C_{ds}$ (\cite{Lilly_LES}) is calculated as
\begin{equation}
    \label{DynCoeff}
    C_{ds} = \frac{1}{2} \frac{\langle M_{ij} \mathscr{L}_{ij} \rangle}{\langle M_{lk} M_{lk} \rangle} \; ,
\end{equation}
where $M_{ij} = \widehat{\Delta} ^ 2 |\widehat{\overline{S}}| \widehat{\overline{S}}_{ij} - \overline{\Delta} ^ 2 \widehat{|\overline{S}| \overline{S}_{ij}}$ and $\langle \, . \, \rangle$ denotes some type of smoothing process such as averaging.


\section{Problem Setup}
\label{section:ProblemSetup}
In this work, the cavitating flow over a stationary NACA66(2)-415 hydrofoil with an elliptical planform is simulated. The geometry of the problem and the working conditions are presented in \S \ref{section:GeoAndWorkCond}, and the strategy employed for grid generation is explained in \S \ref{section:GridGen}, where the derivation of a pressure gradient-based length scale is presented and is used for non-dimensionalization of the mesh size within the tip vortex flow region for mesh requirements specification.

\subsection{Geometry and working conditions}
\label{section:GeoAndWorkCond}
\begin{figure}
    \centering
    \includegraphics[width=0.8\textwidth]{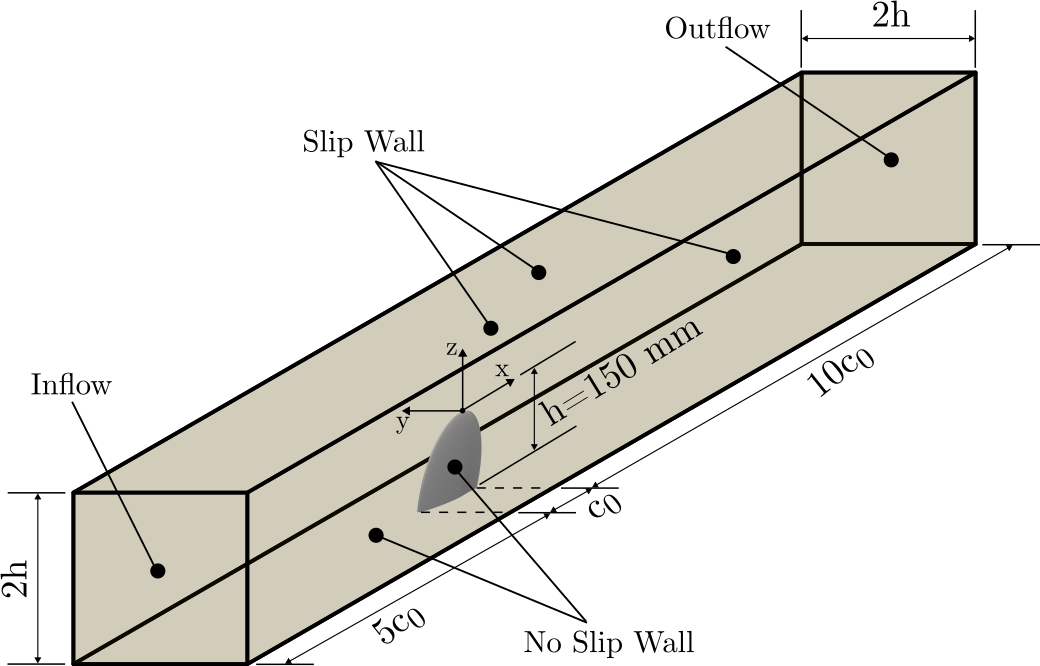}
    \caption{Schematic of the computational domain and boundary conditions}
    \label{fig:Setup}
\end{figure}

The case numerically investigated in this study involves the flow over a stationary NACA66(2)-415 hydrofoil, which is placed within a tunnel with a square cross-section. The geometry and dimensions of the domain and the hydrofoil, depicted in figure \ref{fig:Setup}, are chosen identical to those of the experiments by \cite{Pennings_Experimental} since their experimental data is used herein for validation of the numerical framework.
Due to manufacturing constraints encountered in the experimental study (\cite{Pennings_Experimental}), the foil employed in this investigation is truncated at the trailing edge. The root chord length of the foil after truncation is $c_0 = 0.1256 \; m$ and the half-span of the foil is $h = 0.150 \; \mathrm{m}$. As shown in figure \ref{fig:Setup}, the inflow and outflow boundaries are located $5c_0$ and $10c_0$ in front of and behind the hydrofoil, respectively. The foil's tip is placed at the center of the test section, having a distance of $0.150 \; \mathrm{m}$ from the lateral walls. The origin of the coordinate system is placed at the tip in the simulations, and $x$, $y$, and $z$ directions correspond to the streamwise, lift force, and spanwise directions, respectively, as shown in figure \ref{fig:Setup}.

The inflow velocity and the Reynolds number are kept constant equal to $U_\infty=6.8 \; \mathrm{m/s}$ and $Re= \frac{U_\infty c_0}{\nu} = 8.95 \times 10^5$, respectively, in all of the cases, similar to the experiments. The simulations are carried out at an angle of attack of $\alpha=9 \degree$. Simulations are conducted both for wetted and cavitating conditions. The cavitation number ($\sigma$) is defined based on the outflow pressure as
\begin{equation}
    \label{equation:sigma}
    \sigma = \frac{p_\infty - p_{sat}}{\frac{1}{2} \rho U_{\infty} ^ 2} \; ,
\end{equation}
where $p_{sat}$ and $p_\infty$ are the saturation and the outflow pressure, respectively. In this work, the simulations are carried out at three cavitation numbers $\sigma=1.2$, $1.7$, and $2.6$. The primary emphasis is on the case with a cavitation number of $\sigma=1.7$ due to the occurrence of moderate cavitation in this scenario. The inflow velocity is set to be constant equal to $\mathbf{u}_{in} = (U_\infty, 0, 0)$. The outflow boundary condition is weakly set as constant pressure. The bottom wall of the domain and the hydrofoil surface are set as no-slip walls and other domain walls are of symmetry (slip wall) boundary condition.

In order for the flow to reach statistically stationary conditions in a shorter computation time, the initial condition was obtained from simulations with a coarser grid and a larger time step, and the results were projected onto the desired grid. Using this initial condition, in the first step, the simulations were carried out for $0.2 \, \mathrm{s}$ with a time step of $\Delta t = 10^{-4} \, \mathrm{s}$ in wetted conditions. Subsequently, the time step was decreased to $\Delta t = 10^{-5} \, \mathrm{s}$, which was found to be adequately small for accurate results. Cavitation was started after $0.3 \, \mathrm{s}$ of flow time, and cavitating flow simulations were carried out for $0.5 \, \mathrm{s}$. The results obtained for the last $0.2 \, \mathrm{s}$ of the flow time were used in this work for post-processing and analysis. Due to the computation time limitations, the cases of $\sigma = 2.6$ and $\sigma = 1.2$ were simulated for $0.4 \, \mathrm{s}$ and $0.45 \, \mathrm{s}$, respectively, following the onset of cavitation.

\subsection{Grid generation strategy}
\label{section:GridGen}
It is well-known that due to the high gradients and small scales in tip vortex cavitating flows, adequate mesh resolution in simulating such flows is of critical significance due to the high susceptibility of such flows to numerical diffusion (\cite{2020_Asnaghi_JOE}). Therefore, an adequately fine mesh is required within the tip vortex flow region. Due to the various resolution requirements in different regions of cavitating tip vortex flow, the grid needs to be meticulously designed in order to satisfy mesh requirements while avoiding excessive computational costs. In this work, the grid is designed and generated using Pointwise software. 

\begin{figure}
     \centering
     \begin{subfigure}[t]{0.575\textwidth}
         \centering
         \includegraphics[width=\textwidth]{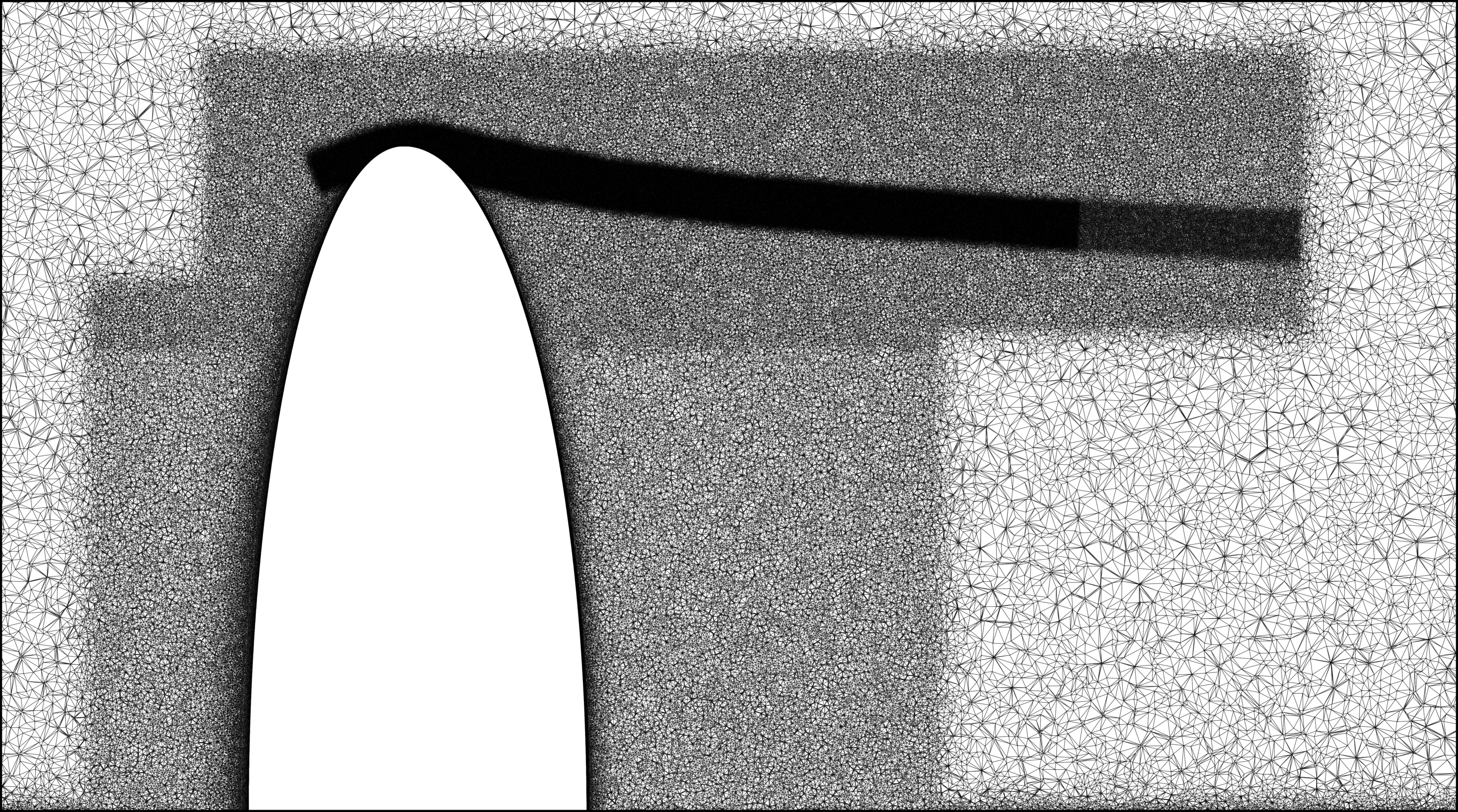}
         \caption{Streamwise distribution}
     \end{subfigure}
     \hfill
     \begin{subfigure}[t]{0.345\textwidth}
         \centering
         \includegraphics[width=\textwidth]{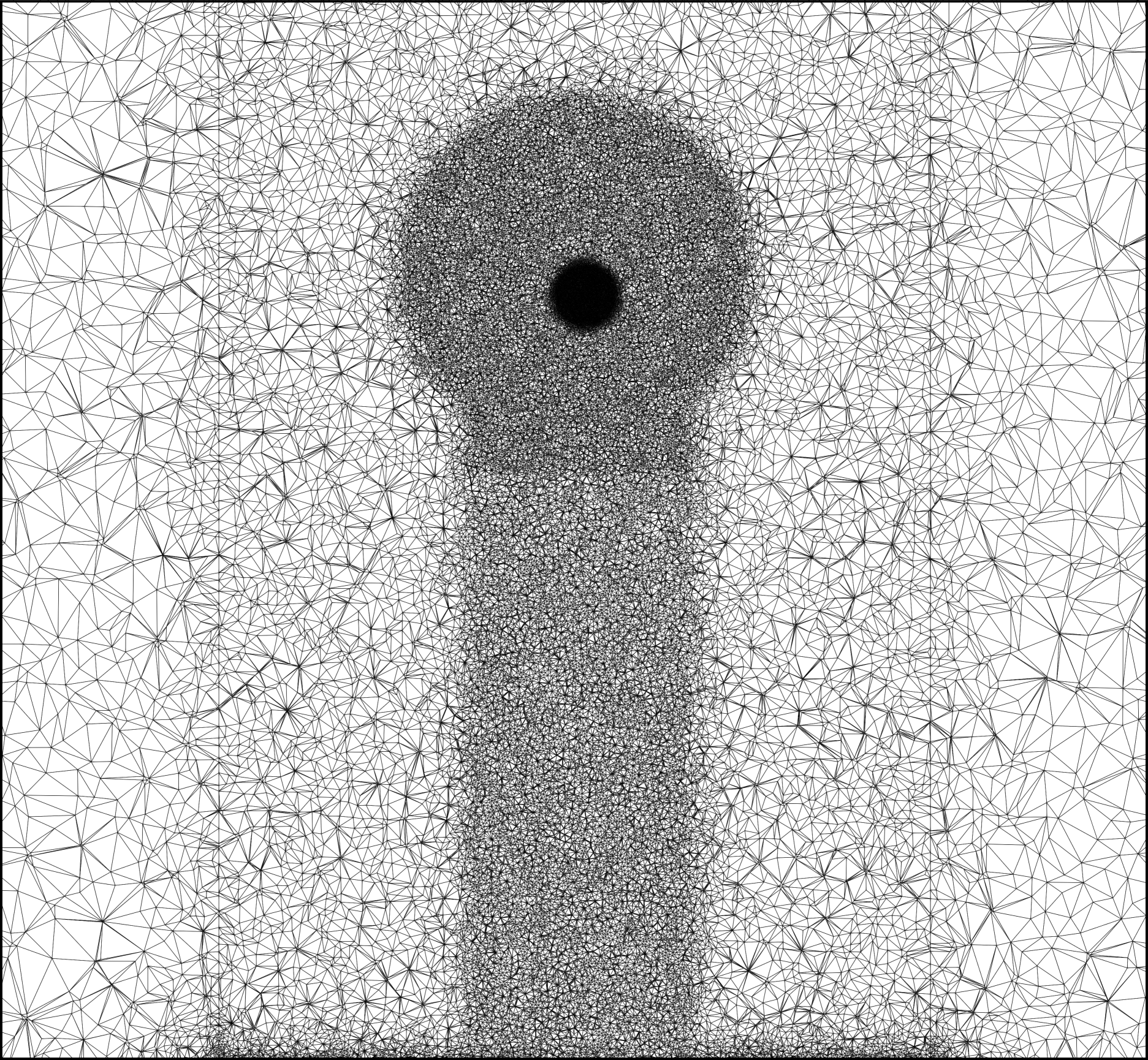}
         \caption{In-plane distribution}
     \end{subfigure}
    \caption{Grid distribution in the streamwise and in-plane directions}
    \label{fig:MeshDist}
\end{figure}
The cross-sectional views of the grid distribution employed in this work in the streamwise and transverse directions are illustrated in figure \ref{fig:MeshDist}. Triangular mesh elements are generated on the hydrofoil surface satisfying the requirements of boundary-layer-resolved LES, i.e., $\{ x^+, z^+ \} < 250$ (\cite{2020_Asnaghi_JOE}). These triangular elements are extruded normal to the hydrofoil surface generating prism layers as the boundary layer mesh. Due to the importance of boundary layer flow in tip vortex cavitation simulation, the height of prism elements is selected to ensure that the dimensionless wall-normal distance ($y^+$) remains below 5. Outside the prism boundary layer, tetrahedral elements are utilized. A refinement region is created enclosing the hydrofoil and the region where flow dynamics are present. A refinement box is added in the wake region to capture the effects of wake dynamics on the tip vortex roll-up process. A cylindrical refinement region with a radius of $r = 30 \; \mathrm{mm}$ is created, starting from a streamwise position slightly upstream of the tip and extending $2 c_0$ downstream of the tip. A smaller refinement region along the path of the tip vortex is created enclosing the tip vortex core to capture the extreme gradients in the tip vortex flow. This region is created by sweeping the tip vortex trajectory by a circle with a radius of $r = 5 \mathrm{mm}$, which is obtained through preliminary simulations with a coarse cylindrical refinement.

The mesh resolution within the tip vortex region is one of the major factors determining the accuracy of tip vortex cavitation simulation. Many researchers have proposed various mesh requirements for simulating tip vortex cavitation (\cite{2020_Asnaghi_JOE, 2013_Ahmad_AIAA}). All of these recommendations are based on the number of grid points across the tip vortex diameter; however, merely considering the number of grid points across the tip vortex diameter does not take all the influencing parameters into account and would not necessarily lead to sufficient resolution. Other characteristics of the tip vortex flow, such as the circulation ($\Gamma$), need to be taken into account as well. There might exist tip vortex flows with the same vortex core diameters but different strengths, and the vortex flow with higher strength, indeed, requires finer mesh resolution. Therefore, a dimensionless mesh resolution within the tip vortex flow region ($\Delta r^*$) is defined herein to account for the effect of the tip vortex strength as well as its core radius.

In the simulation of tip vortex cavitating flows, capturing the radial gradient of pressure $\frac{\partial p}{\partial r}$ is of utmost significance. Therefore, the maximum radial pressure gradient together with the fluid density $\rho$ and kinematic viscosity $\nu$ is used to define a length scale for the tip vortex flow as
\begin{equation}
    \label{lScale}
    \mathcal{L}_\Gamma = \biggl( \frac{\rho \nu ^ 2}{(\frac{\partial p}{\partial r})_{max}} \biggr) ^ {\frac{1}{3}} \; .
\end{equation}
Since the maximum gradient of pressure is not known prior to the simulations and experiments, it needs to be related to the \textit{a priori} known properties of the flow, such as the Reynolds number and the lift coefficient ($C_L$), which is known from the experimental data. In order to relate the maximum radial gradient of pressure to the known parameters of the flow, the tip vortex flow field is assumed to be axisymmetric and the radial velocity is assumed to be zero. In this case, the radial momentum equation simplifies to $\frac{\partial p}{\partial r} = \rho \frac{u_\theta^2}{r}$,
where $r$ is the radial distance from the vortex center, and $u_\theta$ is the azimuthal velocity of the flow. The Rankine vortex model with a circulation of $\Gamma$ and a core radius of $r_v$ is employed to obtain the azimuthal velocity. The Rankine vortex model assumes a solid-body-like rotation within an inner core ($r<r_v$), thereby having a radial velocity distribution of $u_\theta(r) = \frac{\Gamma r}{2 \pi r_v^2}$. The vortex flow outside the rigid core is assumed to be an irrotational vortex ($u_\theta(r) = \frac{\Gamma}{2 \pi r}$).

The maximum radial pressure gradient in a Rankine vortex occurs on the vortex core boundary, i.e., $r=r_v$, where the azimuthal velocity in the Rankine vortex model is $u_\theta = \frac{\Gamma}{2 \pi r_v}$, the substitution of which in the simplified radial momentum equation yields 
\begin{equation}
    \label{dpdRMax-Gamma}
    \left( \frac{\partial p}{\partial r} \right) _{max} = \frac{\rho \Gamma^2}{4 \pi^2 r_v^3}
\end{equation}
for the maximum radial pressure gradient.
Using the maximum radial pressure gradient obtained from equation \ref{dpdRMax-Gamma} in equation \ref{lScale} and rearranging based on the vortex Reynolds number ($Re_\Gamma = \Gamma / \nu$) yields
\begin{equation}
    \label{lengthScalePhysics}
    \mathcal{L}_\Gamma = (2\pi)^\frac{2}{3} r_v Re_\Gamma ^ {-\frac{2}{3}} \; .
\end{equation}
Based on equation \ref{lengthScalePhysics}, the length scale $\mathcal{L}_\Gamma$ scales linearly with the vortex core radius $r_v$ and scales inversely with $Re_\Gamma ^ \frac{2}{3}$, indicating that as the vortex core radius decreases or the vortex strength, represented by the vortex Reynolds number $Re_\Gamma$, increases, the length scale $\mathcal{L}_\Gamma$ decreases, which indicates the need for a finer mesh to be employed in the tip vortex region.

The circulation of the tip vortex $\Gamma$ and the tip vortex core radius $r_v$ are unknown before the numerical analysis or experiments. Therefore, these quantities need to be estimated as well for the calculation of the length scale $\mathcal{L}_\Gamma$. \cite{Fruman} assumed that the local tip vortex circulation and vortex core radius are related to the wing mid-span circulation $\Gamma_0$ and the boundary layer thickness $\delta$, respectively. The mid-span circulation of the hydrofoil $\Gamma_0$ is calculated using the Kutta-Jukowski theorem as $\Gamma _ 0 = \frac{1}{2} C_L U_\infty c_0$.
\cite{1999_Astolfi_EJM} stated that the vortex core radius is directly related to the turbulent boundary layer thickness ($\delta$) of a flat plate with a length of one blade root chord length $c_0$, which can be related to the Reynolds number using the equation $\delta = 0.37 c_0 {Re}^{-0.2}$.
\cite{2015_Pennings_JFM} obtained the coefficients of proportionality $\Gamma/\Gamma_0$ and $r_v/\delta$  to be in the range $0.44<\Gamma/\Gamma_0<0.49$ and $0.56<r_v/\delta<0.61$. In the calculations of \cite{1999_Astolfi_EJM}, these values were in the ranges $0.5<\Gamma/\Gamma_0<0.6$ and $0.8<r_v/\delta<1.0$. These values reveal that since the goal is to find an estimation for the required mesh size within the tip vortex core, the turbulent boundary layer thickness and the mid-span wing circulation can be substituted for the tip vortex core radius and the tip vortex circulation in equation \ref{lengthScalePhysics}, respectively, and all of the coefficients can be omitted since they will merely cause scaling of the results. The pressure gradient-based length scale $\mathcal{L}_\Gamma$ is described in terms of the \textit{a priori} known flow parameters as
\begin{equation}
    \label{lengthScaleFinal}
    \mathcal{L}_\Gamma = \biggl( \frac{16 \pi^2 }{C_L^2 {Re}^{2.6}} \biggr) ^ {\frac{1}{3}}  c_0 \; .
\end{equation}
The length scale $\mathcal{L}_\Gamma$ is employed to non-dimensionalize the mesh element size within the tip vortex region $\Delta r$ as $\Delta r ^ * = \frac{\Delta r}{\mathcal{L}_\Gamma}$.
In this work, simulations are carried out using different values for the non-dimensional mesh resolution $\Delta r ^ *$ within the tip vortex region, and mesh requirements are proposed accordingly in terms of the non-dimensional mesh resolution. The mesh resolution values within the tip vortex flow region and their corresponding non-dimensional values employed in the simulations carried out in this study are presented in table \ref{tab:MeshSizes}, the results of which are presented in \S \ref{section:noncavitating}.

\begin{table}
  \begin{center}
\def~{\hphantom{0}}
  \begin{tabular}{ c  c  c  c } 
    \hline
    Case & Mesh size & $\Delta r^*$ & Number of Nodes \\
    \hline
    I & 0.250 mm & 46 & 10.3 m  \\ 
    II & 0.125 mm & 23 & 13.5 m \\
    III & 0.100 mm & 18.5 & 16.9 m \\
    \hline
  \end{tabular}
  \caption{Simulation cases for mesh resolution effect study}
  \label{tab:MeshSizes}
  \end{center}
\end{table}



\section{Results and discussion}
\label{section:results}
In this section, the results obtained from the simulations carried out using our finite element solver, the details of which were discussed in \S \ref{section:ComputationalFramework}, are presented. In \S \ref{section:noncavitating}, the results of non-cavitating tip vortex flow simulations are compared with the experimental data (\cite{Pennings_Experimental}), and the effect of mesh resolution on the accuracy of the simulations is investigated. The tip vortex cavitating flow simulation results are showcased in \S \ref{section:TVCOscillations} and assessed with the semi-analytical solution by \cite{Bosschers_thesis_2018}. The breathing mode of oscillation is further investigated at different cavitation numbers in \S \ref{section:Reff}, and the pressure fluctuations within the domain are analyzed in \S \ref{section:PressureFluctuations} to ascertain the correlations between the breathing mode oscillations of the tip vortex cavity and the pressure fluctuations within the domain. 

\subsection{Non-cavitating flow}
\label{section:noncavitating}
To study the effect of mesh resolution on the wetted tip vortex flow simulation, and to validate the capability of the presented numerical framework in simulating such flows, the axial ($x$ direction) and in-plane (parallel with $yz$ plane) velocity distribution at three different downstream locations ($x/c_0 = 0.5, \, 0.75, \, 1.14$) are compared qualitatively with the experimental data, and quantitative comparison is carried out for the time-averaged lift coefficient and the in-plane velocity distribution at the downstream location $x/c_0 = 1.14$. 
Due to the effect of the lift coefficient on the tip vortex flow field, both the lift coefficient and the velocity profiles need to be accurately predicted in the simulations in order for the tip vortex cavitating flow simulations to have adequate accuracy.

\begin{figure}
     \centering
     \begin{subfigure}[t]{0.4\textwidth}
        \centering
         \includegraphics[width=\textwidth]{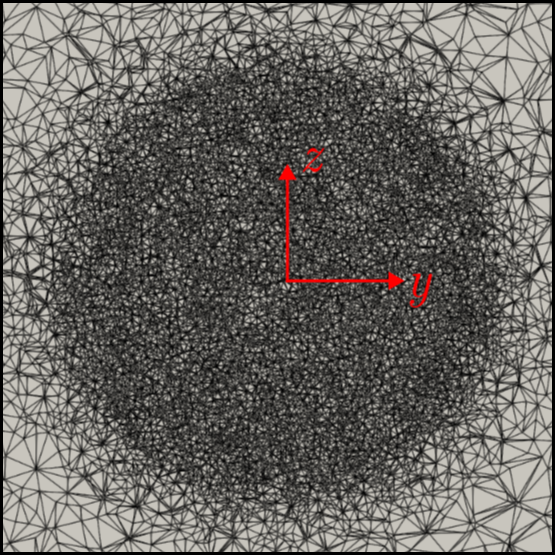}
         \caption{Triangular mesh on a 2D plane orthogonal to the free-stream flow direction obtained from interpolation from the 3D mesh}
         \label{fig:InterpMesh:2D}
     \end{subfigure}
     \hfill
     \begin{subfigure}[t]{0.422\textwidth}
        \centering
        \includegraphics[width=\textwidth]{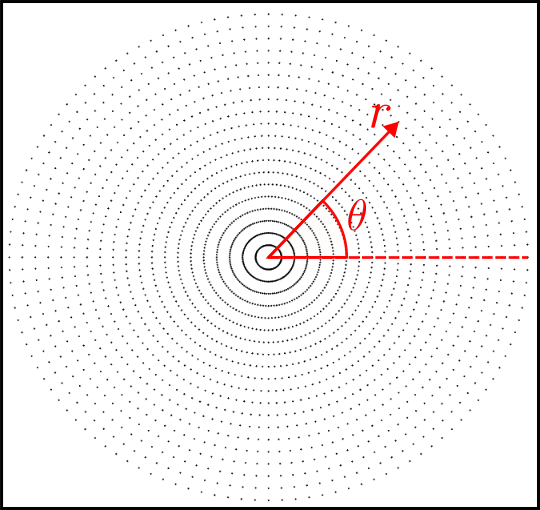}
        \caption{Polar grid points generated around the tip vortex center for interpolation of the data from the 2D cross-section of the flow}
        \label{fig:InterpMesh:Polar}
     \end{subfigure}
    \caption{Projected 2D triangular mesh and polar mesh used during the data processing for comparison with the experimental data}
    \label{fig:InterpMesh}
\end{figure}
\begin{figure}
     \centering
     \begin{subfigure}{\textwidth}
         \includegraphics[width=\textwidth]{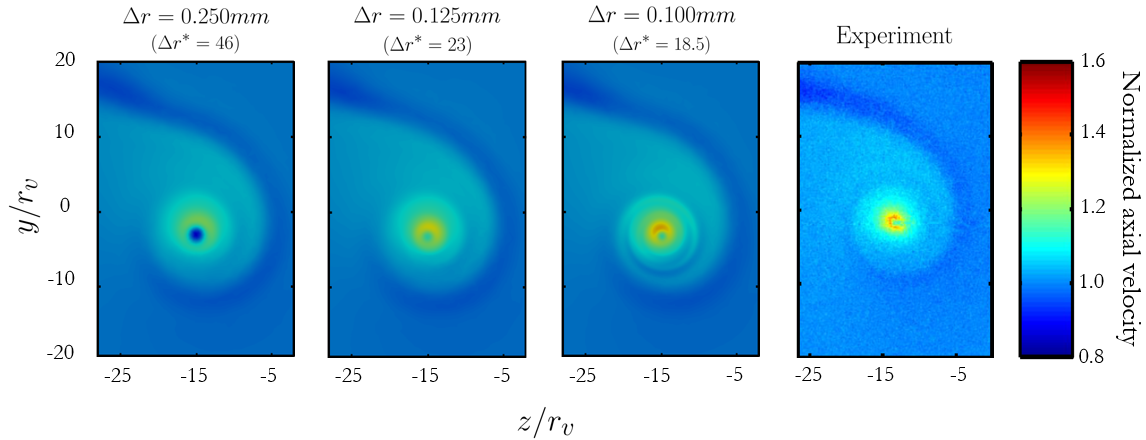}
         \caption{Non-dimensional Axial Velocity}
         \label{fig:MeshEffectContour:Axial}
     \end{subfigure}
     \hfill
     \begin{subfigure}{\textwidth}
         \includegraphics[width=\textwidth]{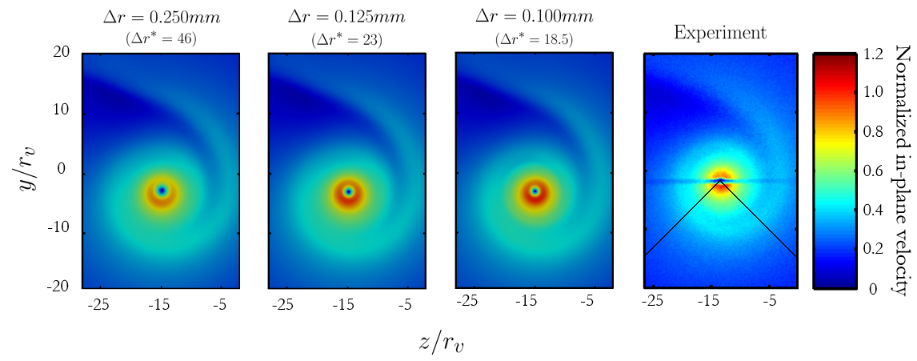}
         \caption{Non-dimensional In-plane Velocity}
         \label{fig:MeshEffectContour:Inplane}
     \end{subfigure}
    \caption{Comparison of non-dimensional velocity distribution within the tip vortex region at downstream location of $x/c_0 = 1.14$ obtained from simulations with different mesh sizes with the experimental measurements}
    \label{fig:MeshEffectContour}
\end{figure}

The data obtained from the simulations need to be processed prior to comparison with the experimental data. First, the vortex core center is identified at every time step assuming that the minimum pressure within the vortex core occurs at the vortex center. In the next step, the results obtained from 3D simulations are projected onto a 2D cross-section of the flow orthogonal to the streamwise direction (parallel with the $yz$ plane). The mesh projected onto the 2D plane is illustrated in figure \ref{fig:InterpMesh:2D}. In the next step, a new polar coordinate system is defined with the vortex core center as the origin, and the data is interpolated from the 2D plane of interest onto a polar mesh, depicted in \ref{fig:InterpMesh:Polar}, around the tip vortex core center using cubic spline interpolation method. This process is carried out for different flow times, and then the quantities obtained for various flow times on the vortex-based polar coordinate system are averaged after aligning the tip vortex center obtained for all flow times used in averaging. Utilizing this method of post-processing eliminates the effect of vortex wandering on time-averaging. For the quantitative comparison of azimuthal velocity at the downstream location $x/c_0 = 1.14$ with the experimental measurements, the in-plane velocity values are spatially averaged on $90\degree$ arcs between the two perpendicular black lines illustrated in \ref{fig:MeshEffectContour:Inplane}, same as the averaging method used in \cite{Pennings_Experimental} since the asymmetry of the tip vortex flow field at this location necessitates the utilization of the same averaging method for the sake of accuracy.

\begin{figure}
    \centering
    \includegraphics[width=0.6\textwidth]{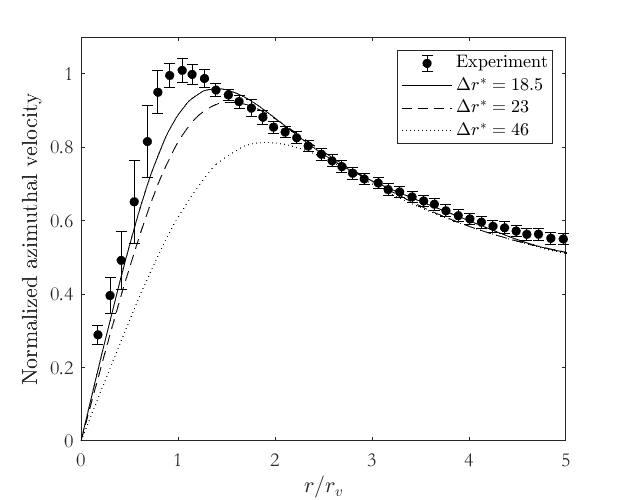}
    \caption{Comparison of normalized azimuthal velocity profile at $x/c_0 = 1.14$ obtained from simulations using various grid sizes with experimental data}
    \label{fig:MeshEffectPlot}
\end{figure}

As a quantitative comparison, the time- and contour-averaged azimuthal velocity profiles predicted by the simulations employing various mesh sizes within the tip vortex region at the cross-section $x/c_0 = 1.14$ are shown in figure \ref{fig:MeshEffectPlot}. This plot shows that accurately capturing the velocity distribution of tip vortex requires extremely high mesh resolution within this region, and utilization of inadequate resolution leads to excessive dissipation and underprediction of the maximum azimuthal velocity and its gradient. Moreover, employing adequately small grid elements within the tip vortex region is necessary for the correct prediction of the tip vortex core radius. The results presented in figure \ref{fig:MeshEffectPlot} indicate that the results obtained from the simulation with $\Delta r ^ * = 46$ are excessively dissipated compared to the experimental data. Refining the grid to $\Delta r ^ * = 23$ led to the overall correct capturing of the velocity distribution, but the tip vortex flow field, in this case, is slightly dissipated. The utilization of a grid with $\Delta r ^ * = 18.5$ led to accurate prediction of the tip vortex flow field.

Based on the qualitative and quantitative comparisons of the predicted axial and azimuthal velocity field presented herein, it can be concluded that the non-dimensional grid size $\Delta r ^ *$ requires to be chosen below 20 in order for the simulations to accurately capture the dynamics of tip vortex flow. This guideline can serve as a general mesh resolution criterion for future works on tip vortex flow simulation.
Furthermore, the agreement between the results obtained from the simulation using a non-dimensional mesh resolution of $\Delta r ^ * = 18.5$ in the tip vortex region reveals the capability of the developed finite element-based numerical framework in simulating non-cavitating tip vortex flows.

\subsection{Overview of TVC Oscillatory Dynamics}
\label{section:TVCOscillations}
\begin{figure}
    \centering
    \includegraphics[width=\textwidth]{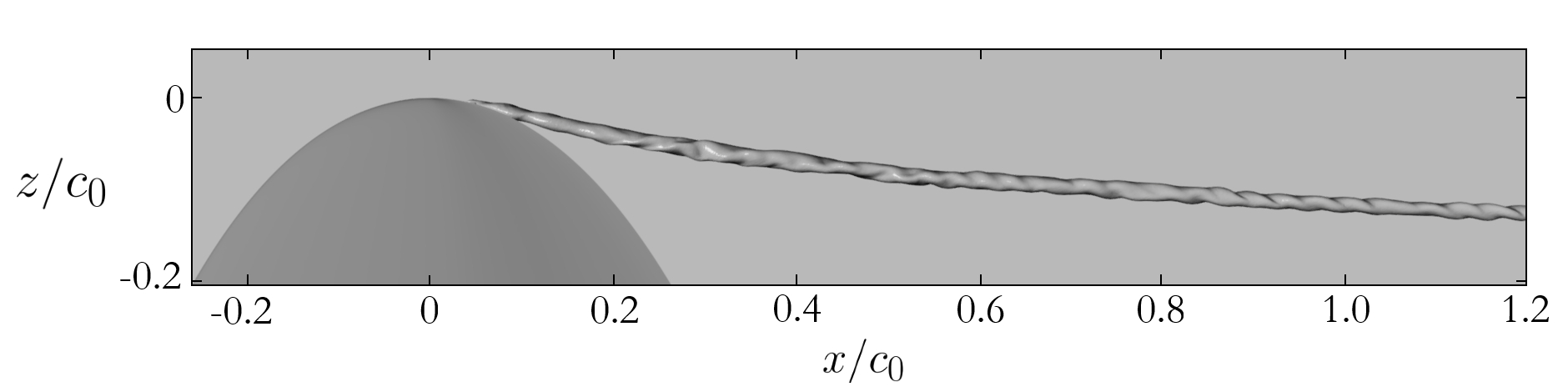}
    \caption{Tip vortex cavity surface observed in simulations with a cavitation number of $\sigma = 1.7$}
    \label{fig:TVCContour}
\end{figure}

Similar to the experimental observations (\cite{2015_Pennings_JFM}), the tip vortex cavity observed in the simulations carried out in this work exhibits complex surface oscillations as shown in figure \ref{fig:TVCContour}. As explained in \S \ref{section:Introduction}, these surface oscillations have been found to be one of the sources of TVC noise. Therefore, investigating the physics of these oscillations and deformations, especially the breathing mode of oscillation, is crucial to understanding the contributions of TVC to the URN. In this section, an overview of the oscillations of the tip vortex cavity surface is presented for the case of $\sigma=1.7$ using a method similar to what is commonly employed in experimental studies (\cite{2015_Pennings_JFM}) to extract various modes of cavity surface oscillations, which uses the diameter of the cavity as viewed from different sides to distinguish between different modes of oscillation. Furthermore, the results obtained using this method are compared with the analytical solution proposed by \cite{Bosschers_thesis_2018}.



\begin{figure}
\centering
    \begin{subfigure}[b]{\textwidth}
        \includegraphics[width=\textwidth]{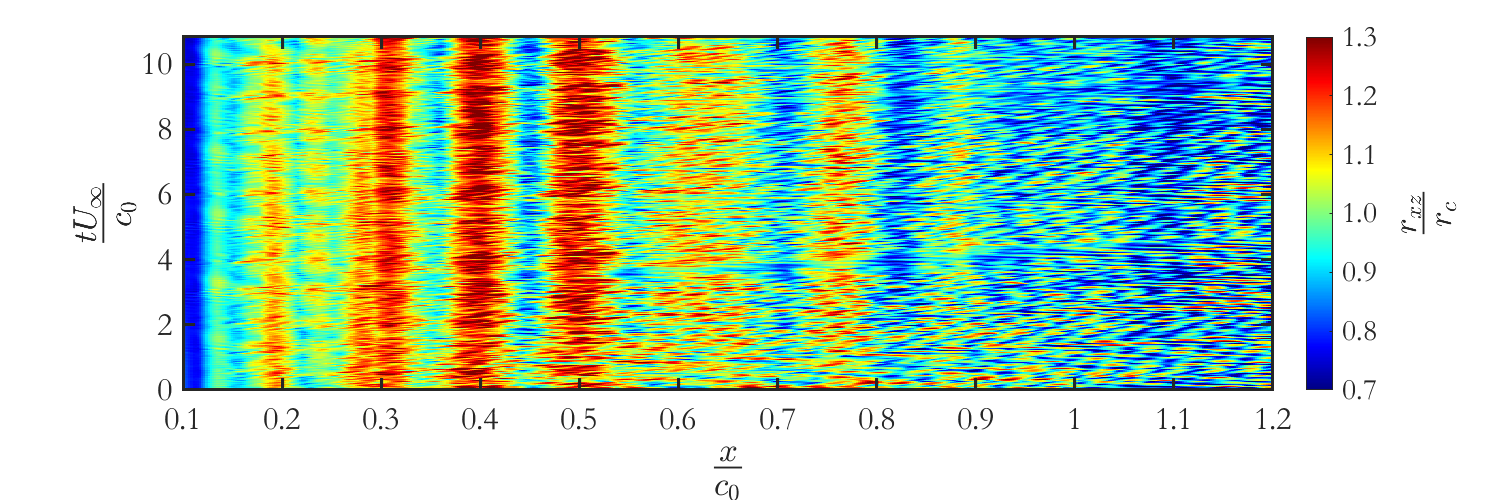}
        \caption{$xz$-plane (side view)}
        \label{fig:XZViewDiameter}
    \end{subfigure}
    \hfill
    \begin{subfigure}[b]{\textwidth}
        \includegraphics[width=\textwidth]{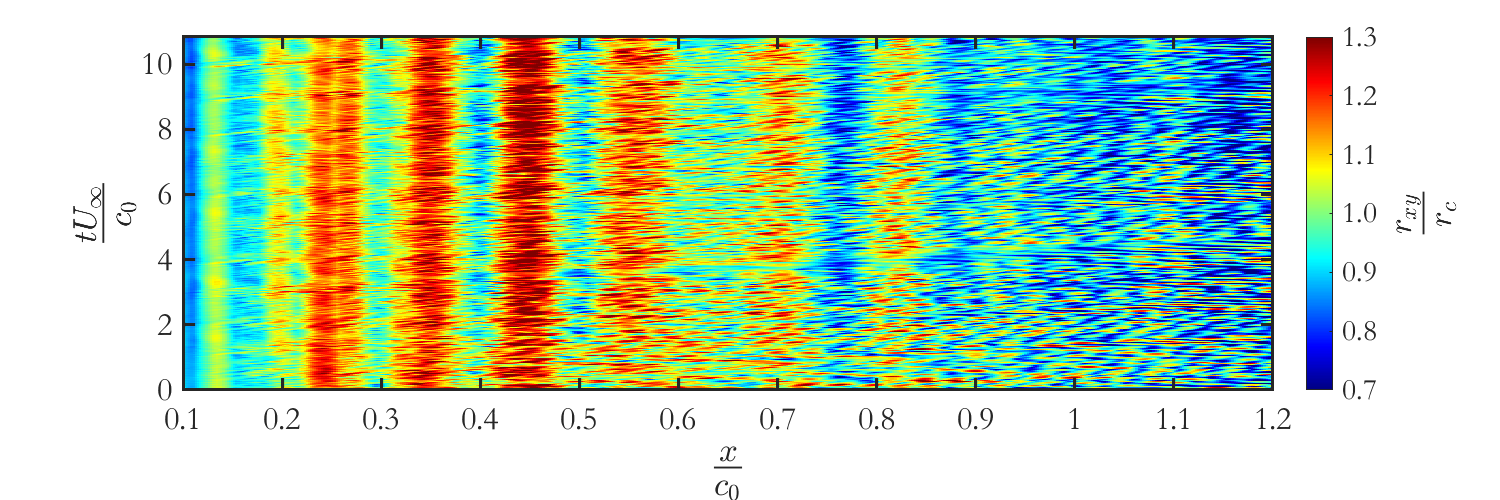}
        \caption{$xy$-plane (top view)}
        \label{fig:XYViewDiameter}
    \end{subfigure}
    \caption{Spatial-temporal variation of the non-dimensional radius observed at $\sigma = 1.7$ from: (a) side ($xz$-plane) and (b) top ($xy$-plane) views.}
    \label{fig:TwoViewsDiameter}
\end{figure}

In order to extract various modes of the cavity surface oscillations, illustrated in figure \ref{fig:CavityModeShapes}, from the complex cavity surface similar to the tip vortex cavity illustrated in figure \ref{fig:TVCContour}, the tip vortex cavity diameter observed from different views is employed in experimental studies (\cite{2015_Pennings_JFM, Bosschers_thesis_2018}). This method can be utilized in numerical studies to attain an overview of the oscillations captured in the simulations, which can be used as a basis for more complex post-processing of the results. The non-dimensional tip vortex cavity radius observed from the side ($xz$ view) and top ($xy$ view) of the domain is presented in figure \ref{fig:TwoViewsDiameter}, which clearly indicates the presence of cavity surface oscillations in the results of the simulations. The radius used for non-dimensionalizing the data throughout this work, i.e., $r_c$, is the average of the spatially- and temporally-averaged radius observed from the top and side views in the region of interest ($0.1 < x/c_0 < 1.2$).

Figure \ref{fig:TwoViewsDiameter} shows that the cavity surface exhibits a relatively stationary behavior from the tip until $x/c_0 \sim 0.5$ (formation region), and transitions to a more dynamic behavior further downstream (developed region). Comparing figures \ref{fig:XZViewDiameter} and \ref{fig:XYViewDiameter} reveals that the spatial diameter variations of the roll-up region of the cavity ($0.1 < x/c_0 < 0.5$) observed from these two views are out-of-phase, which indicates the presence of a nearly stationary twisting shape (double-helical mode) in this region of the cavity similar to the observations of \cite{2023_Ye_JFM}.
In order to extract the cavity oscillation modes from the side-view and top-view diameter, similar to the method employed by \cite{Bosschers_thesis_2018}, the cross-power spectral density (CPSD) is calculated from the two-dimensional Fast Fourier Transform (FFT) of the data obtained from the two views using the equation
\begin{equation}
    S(k_x, f) = 120 + 10 \log_{10} \left[ \frac{G_{top}(k_x, f) G^*_{side}(k_x, f)}{r_c^2} \right] \; ,
    \label{eq:CPSD}
\end{equation}
where $k_x$ denotes the streamwise wavenumber, $G(k_x, f)$ represents the 2D FFT of the radius data observed from the corresponding view, and $G^*(k_x, f)$ is its complex conjugate. 

\begin{figure}
\centering
\begin{subfigure}[t]{.48\textwidth}
    \centering
    \includegraphics[width=\textwidth]{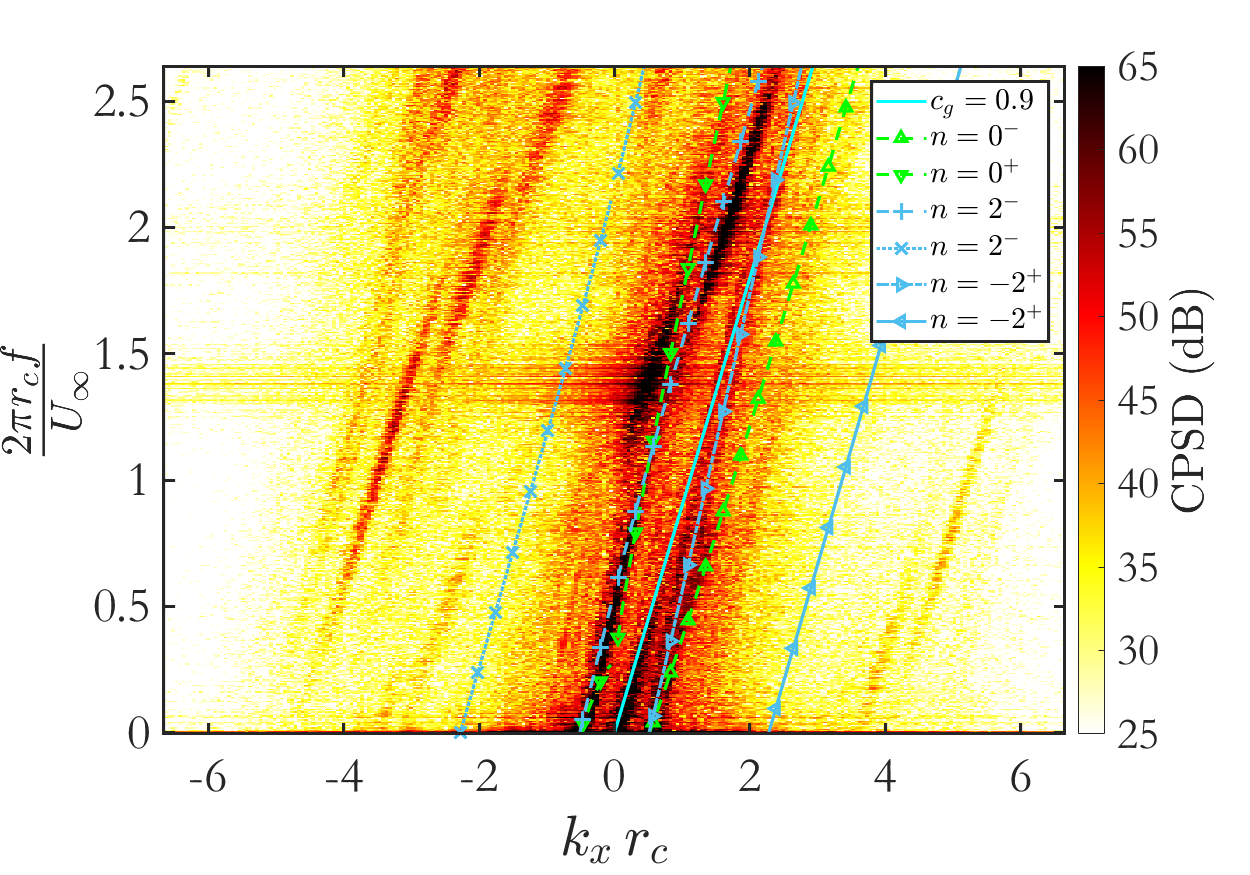}
    \caption{Cross-power spectral density obtained from Equation \ref{eq:CPSD}}
    \label{fig:CPSD}
\end{subfigure}%
~
\begin{subfigure}[t]{.48\textwidth} 
    \centering
    \includegraphics[width=\textwidth]{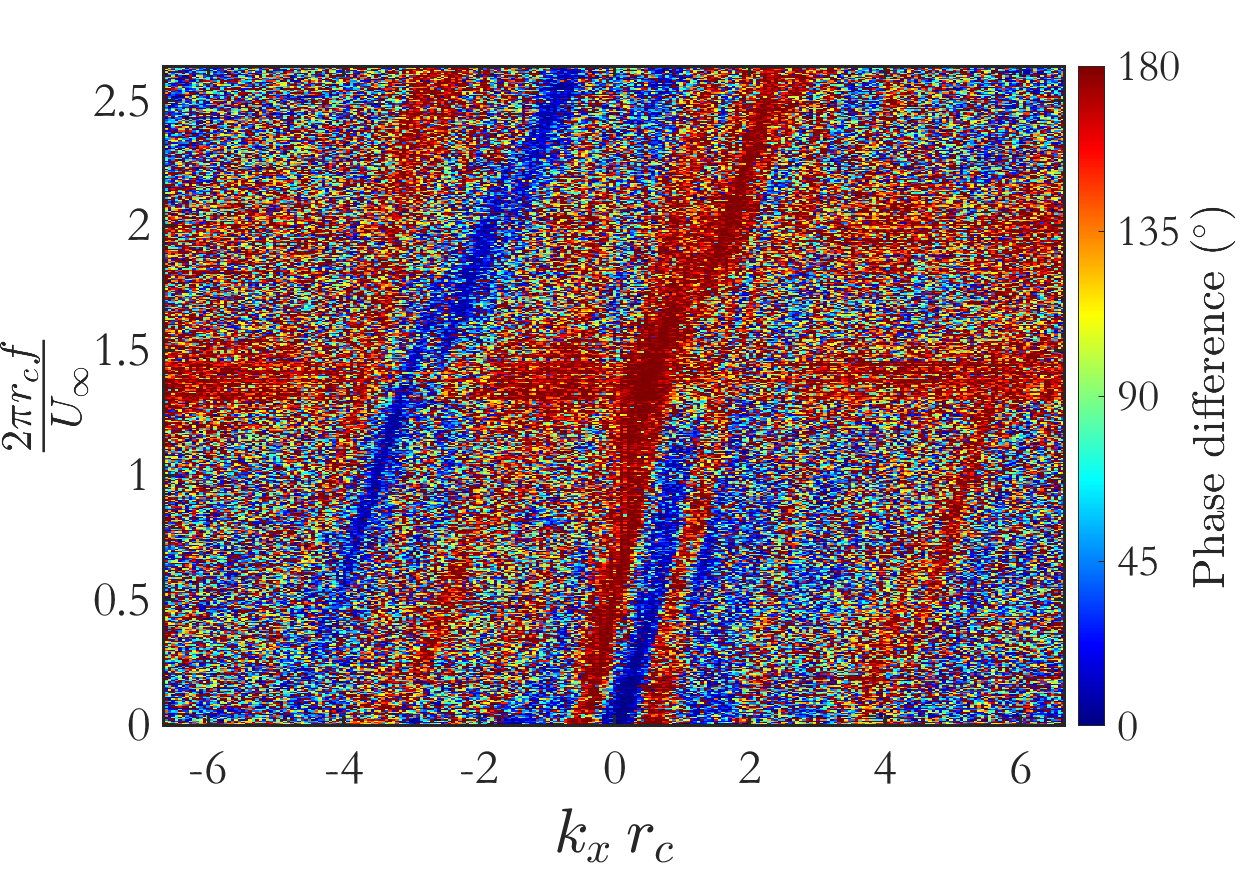}
    \caption{Phase difference between 2D FFT of cavity diameter viewed from the side and top views}
    \label{fig:PhaseDiff}
\end{subfigure}
\caption{Cavity oscillatory mode extraction using two orthogonal views of the cavity}
\end{figure}

The wavenumber-frequency dependence of $S(k_x, f)$ is depicted in figure \ref{fig:CPSD}. Various modes of cavity oscillation manifest themselves as curves with higher $S(k_x, f)$ values as shown in figure \ref{fig:CPSD}. The phase difference between the 2D FFT of the radius data obtained from the side and top views is also required in order to distinguish between the breathing and double helical modes. The phase difference, illustrated in figure \ref{fig:PhaseDiff}, is equal to $180 ^ {\circ}$ in double helical mode and $0 ^ {\circ}$ in breathing mode oscillations. The results depicted in these figures clearly indicate the presence of the breathing and the double helical modes.

The analytical solution proposed by \cite{Bosschers_thesis_2018} is also plotted in figure \ref{fig:CPSD} for comparison with the results obtained from the simulations in this work. The analytical solution for the dispersion relation of the cavity surface waves is
\begin{equation}
    \frac{2 \pi r_c f ^ {\pm}}{U_\infty} = \tilde{U}_x k_x r_c + \tilde{U}_\theta n \pm \sqrt{K_\sigma} \sqrt{\frac{- | k_x r_c | \, K'_n(| k_x r_c |)}{K_n(| k_x r_c |)}} T_\omega \; ,
    \label{eq:AnalyticalBosschers}
\end{equation}
where $\tilde{U}_x$ and $\tilde{U}_\theta$ are the non-dimensional axial and azimuthal velocities, respectively. $K_n$ and $K'_n$ denote the modified Bessel function of the second kind and its first derivative, respectively. $K_\sigma$ is a non-dimensional stiffness coefficient, and $T_\omega$ represents the effect of surface tension, which is neglected in this work ($T_\omega = 1$). The values $\tilde{U}_x = 1.05$, $\tilde{U}_\theta = 0.68$ and $K_\sigma = 0.32$ are used herein.

The results illustrated in figure \ref{fig:CPSD} indicate that all of the four double helical mode oscillations ($n=\pm 2 ^ {\pm}$) are captured in the simulations, and these modes of cavity oscillation observed in the simulations agree reasonably well with the analytical results. The modes $n = 2^{+}$ and $n = -2^{-}$ observed in the results obtained from simulations deviate from the analytical results, which might be resolved if the parameters in equation \ref{eq:AnalyticalBosschers} are adjusted further. Moreover, some high-CPSD curves with $0^{\circ}$ phase difference can be observed in figure \ref{fig:CPSD} that are not predicted by the analytical model.

Another dominant stripe observed in figure \ref{fig:CPSD}, which exhibits a phase difference of $0^{\circ}$ between the two views, shows the convection of the disturbances on the cavity surface. The convection group velocity ($c_g$) in this case is found to be $c_g = 0.9 U_\infty$, which is similar to the value used by \cite{Bosschers_thesis_2018}, i.e., $c_g = 0.95 U_\infty$.
In addition, figure \ref{fig:CPSD} shows that the wavenumber-frequency lines obtained for the breathing and double helical modes are close in this case, and the dominance of the double-helical mode oscillations in this case hinders the detection of the breathing mode oscillations. Only traces of the breathing mode oscillations can be seen in the phase difference between the two views shown in figure \ref{fig:PhaseDiff}, which demonstrates that a more effective method is required for the extraction of the breathing mode from the cavity oscillations.
These results indicate that the developed numerical framework in this work is capable of accurately capturing the tip vortex cavity surface oscillations, which allows further investigation of the cavity surface oscillations and their contribution to the radiated noise.

\subsection{Breathing Mode Oscillations}
\label{section:Reff}
A major proportion of the TVC noise, especially in low and medium frequencies, is attributed to the mode of oscillation responsible for cavity volume variations; i.e., the breathing mode (\cite{Wang_POF_2023}). Moreover, the center frequency of the broadband hump observed in the URN spectrum is hypothesized to be related to the tip vortex cavity resonance frequency (\cite{Bosschers_thesis_2018}), and experimental evidence has been provided by \cite{2015_Pennings_JFM} showing that this resonance frequency is the frequency of zero group velocity of the breathing mode. These connections linking the breathing mode to the TVC noise necessitate a comprehensive investigation of this mode of cavity oscillation. Therefore, in this section, this mode of oscillation is investigated in detail for three cases with different cavitation numbers. First, in \S \ref{section:ReffExtraction}, a new method is developed for extracting the breathing mode from the cavity surface obtained from the simulations by defining an effective radius for the tip vortex cavity. The temporally- and spatially-averaged effective radius of the cavity are analyzed in \S \ref{section:TemporalAvgReff} and \S \ref{section:SpatialAvgReff}, respectively. \S \ref{section:Reff2DFFT} focuses on the spatial and temporal development of the effective radius delving into the characteristics of the local breathing mode oscillations. Proper Orthogonal Decomposition (POD) is performed on the effective radius data in \S \ref{section:ReffPOD} to further analyze the streamwise development of the tip vortex cavity's volume.

\subsubsection{Breathing Mode Extraction}
\label{section:ReffExtraction}

\begin{figure}
\centering
\begin{minipage}[t]{.5\textwidth}
    \centering
    \includegraphics[width=\textwidth]{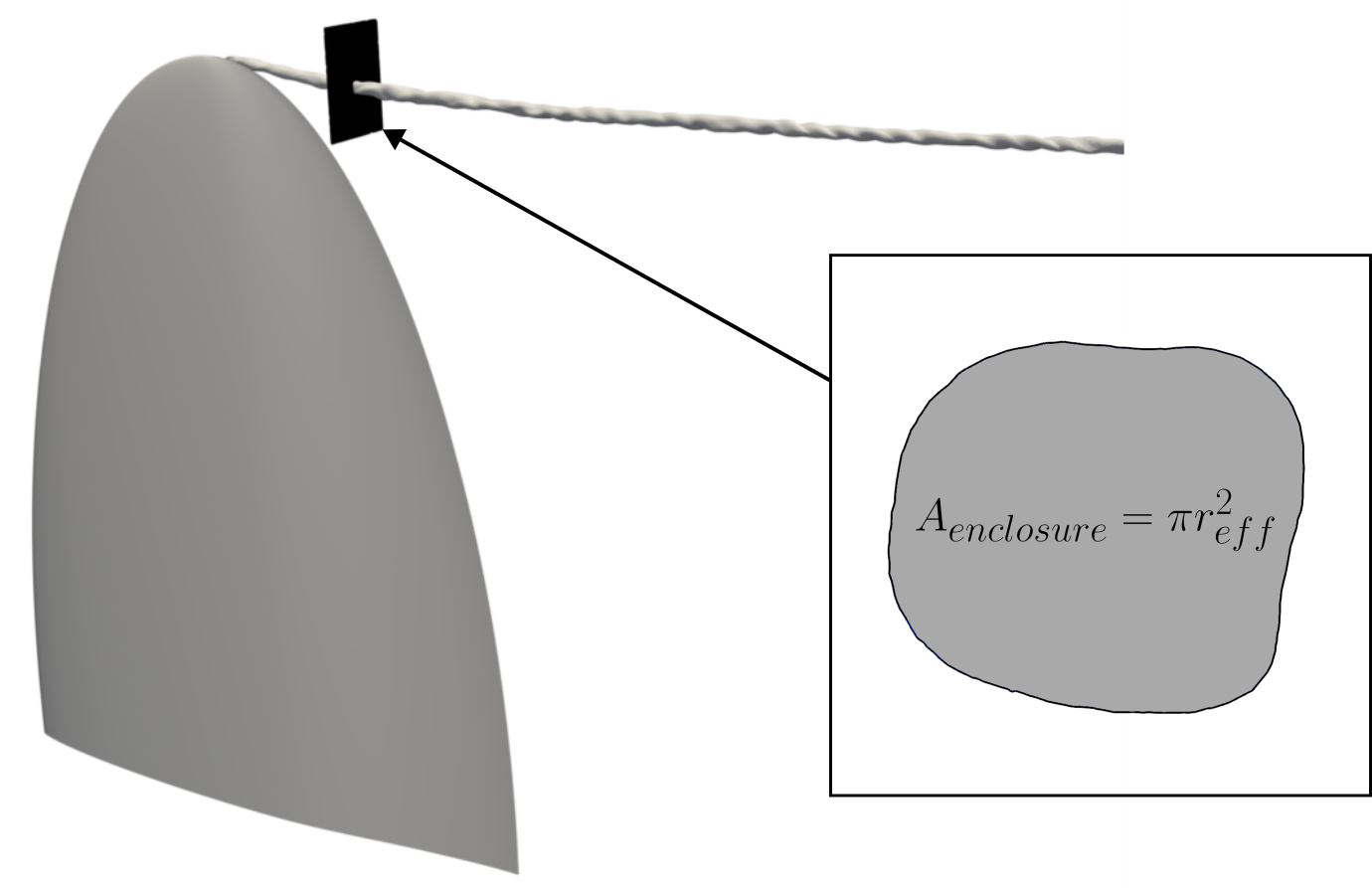}
    \caption{Calculation of the effective radius using the tip vortex cavity surface data}
    \label{fig:NonEllipticalCross}
\end{minipage}%
\hfill
\begin{minipage}[t]{.48\textwidth} 
    \centering
    \includegraphics[width=\textwidth]{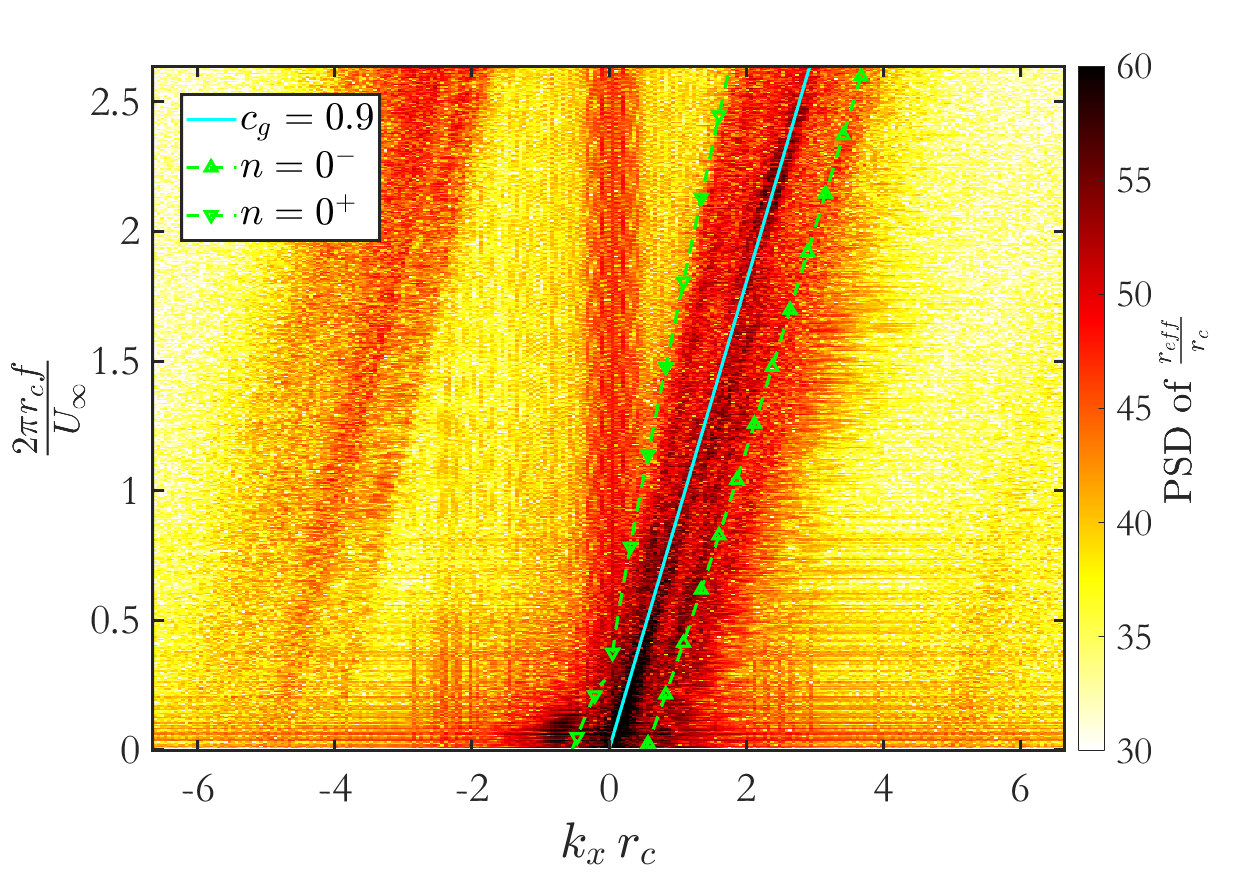}
    \caption{Frequency-wavenumber diagram of the effective radius data ($\sigma = 1.7$)}
    \label{fig:Reff2DFFT}
\end{minipage}
\end{figure}

In order to examine the breathing mode oscillations in detail, a method is required for extracting this mode of cavity oscillation. \cite{Klapwijk_JOE_2022} fitted an ellipse onto the transverse cross-section of the cavity interface, and defined an effective radius as $r_{eff}=\sqrt{ab}$, where $a$ and $b$ are the semi-major and semi-minor axes of the fitted ellipse, respectively. However, as explained by \cite{Wang_POF_2023}, the method used in \cite{Klapwijk_JOE_2022} might lead to loss of deformation information when the cross-section is non-elliptical. Such a highly non-elliptical transverse cross-section of the cavity was also observed in our simulations as shown in figure \ref{fig:NonEllipticalCross}.

Since the breathing mode is the only cavity oscillation mode responsible for the cavity volume variations, the idea of employing the cavity's cross-sectional surface area for calculating an effective radius is a sound approach. Based on this approach, first, the transverse cross-section of the cavity at different downstream locations is extracted, the area enclosed by which is then calculated. Subsequently, an effective radius is calculated based on the area enclosed by the cavity cross-section as $A_{enclosure} = \pi r_{eff}^2$ as depicted in figure \ref{fig:NonEllipticalCross}. The effective radius obtained using this method only contains the volume variation information, and other forms of oscillation are excluded from this quantity. Therefore, this parameter can be employed to investigate breathing mode oscillation in the absence of the interference of other modes.

The two-dimensional Fourier transform of the effective radius obtained using this method for the case with a cavitation number of $\sigma = 1.7$ is illustrated in figure \ref{fig:Reff2DFFT}, which displays the wavenumber-frequency dependency of the breathing mode cavity oscillations together with the analytical solution. Based on this figure, the breathing mode oscillations obtained from performing the extraction method on the simulation data agree well with the analytical solution, which shows that this extraction method together with the numerical framework employed in this work is able to accurately capture the breathing mode oscillations.


\subsubsection{Temporally-averaged Effective Radius}
\label{section:TemporalAvgReff}

The tip vortex cavity surface can be decomposed into a mean cavity surface and time-dependent cavity surface oscillations. The formation of the mean cavity interface is related to the overall roll-up process of the tip vortex. In this section, the impact of the roll-up process on the mean cavity volume is investigated by analyzing the time-averaged effective radius. The goal here is to gain insight into the evolution of the tip vortex cavity from its inception to decay.

\begin{figure}
    \centering
    \includegraphics[width=0.8\textwidth]{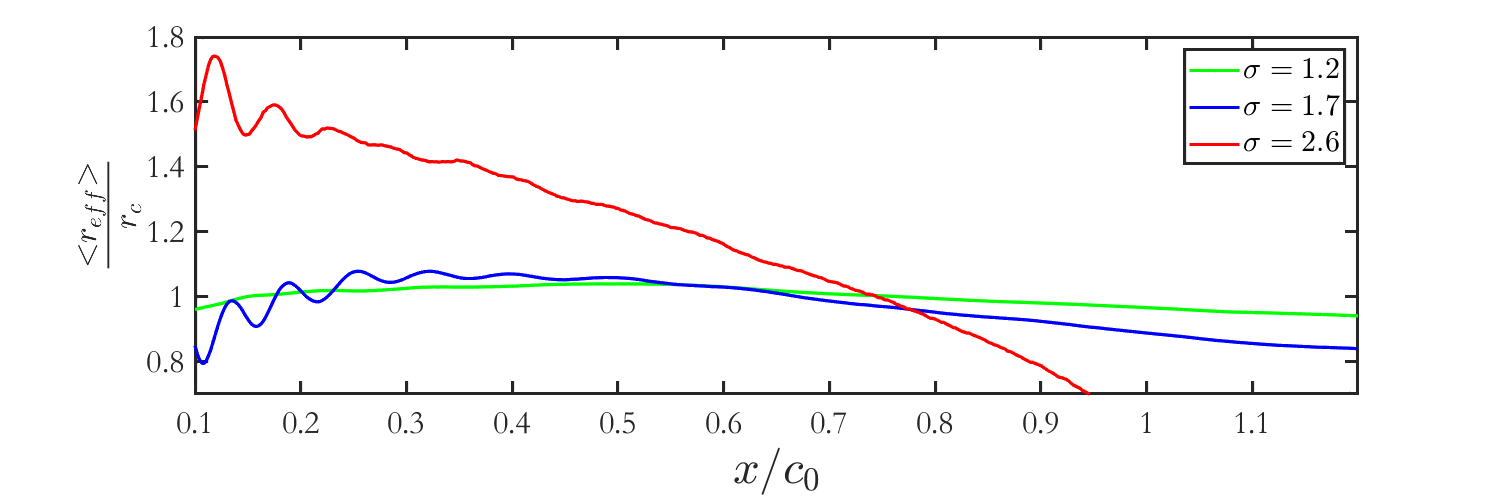}
    \caption{Streamwise evolution of the temporally-averaged effective radius in different cavitation numbers}
    \label{fig:TimeAvgReff}
\end{figure}

Figure \ref{fig:TimeAvgReff} shows the temporally-averaged effective radius, denoted by $<r_{eff}>$ and non-dimensionalized by $r_c$, with respect to the streamwise location ($x$) non-dimensionalized by the root chord length ($c_0$) at three cavitation numbers. According to the streamwise development of the temporally-averaged effective radius in these cases, the cavity undergoes a region of growth, where the cavity is forming and the cavity effective radius has a growing trend overall, followed by a decay region.

A comparison of different cavitation numbers indicates that the growth region shrinks as the cavitation number increases. In the case with $\sigma = 2.6$, the decay starts approximately at $x/c_0 = 0.12$. In the case with $\sigma = 1.7$, the growth region extends until $x/c_0 = 0.4$, and when $\sigma = 1.2$, the growth continues further until $x/c_0 = 0.53$.
Furthermore, figure \ref{fig:TimeAvgReff} reveals that the decay occurs at a greater slope in higher cavitation numbers. The most sustained cavity is observed in the case with $\sigma = 1.2$, where the slope of decay ($\frac{dr^*}{dx^*}$, where $r^* = \frac{r_{eff}}{r_c}$ and $x^* = \frac{x}{c_0}$) is approximately $0.146$. This slope of decay in the cases of $\sigma = 1.7$ and $\sigma = 2.6$ increases to $0.288$ and $1.027$, respectively. 

An interesting feature of the temporally-averaged effective radius observed in figure \ref{fig:TimeAvgReff} is the spatial oscillations observed in the growth region. In the case with $\sigma = 1.7$, the cavity volume exhibits periodic spatial oscillation with the downstream location in the growth region ($x/c_0 < 0.4$); however, in the decay region, these spatial oscillations of the cavity volume vanish. When $\sigma = 2.6$, such oscillations are present until $x/c_0 = 0.23$, which can be attributed to the smaller growth region in this case compared to $\sigma = 1.7$. On the other hand, as it is evident in this figure, the spatial volumetric oscillations in the formation region vanish when the cavitation number is lowered to $\sigma = 1.2$.

\begin{figure}
    \centering
    \includegraphics[width=0.5\textwidth]{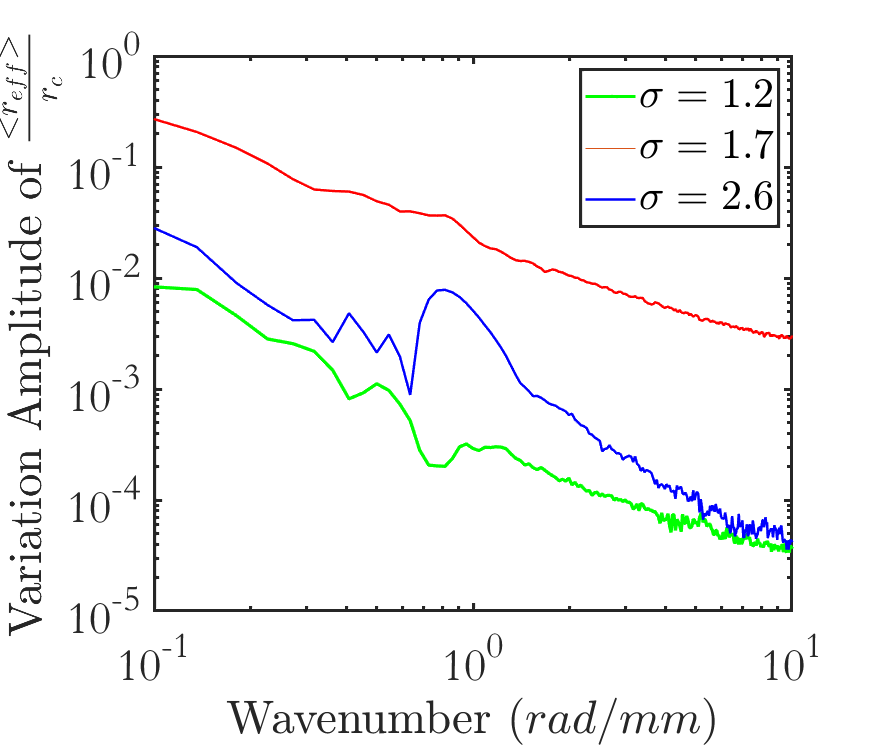}
    \captionof{figure}{Fourier transform of the non-dimensional temporally-averaged effective radius variations} 
    \label{fig:TimeAvgReffFFT}
\end{figure}

The spatial variations of the temporally-averaged effective radius in the cavity growth region are further investigated by calculating their Fourier transform, which is illustrated in figure \ref{fig:TimeAvgReffFFT}. The large-amplitude low-wavenumber components ($\sim 0.1 \, \mathrm{rad/mm}$) correspond to the overall growth and decay of the cavity. An interesting feature observed in the case of $\sigma = 1.7$ is the hump within the wavenumber range of $0.68  \, \mathrm{rad/mm} <k< 1.09  \, \mathrm{rad/mm}$, which corresponds to a wavelength range of $0.074 \, c_0 > \lambda > 0.046 \, c_0$. The center wavenumber of this hump is $0.82 \, \mathrm{rad/mm}$, corresponding to a wavelength of $\lambda = 0.061 \, c_0$. A local maximum at this wavenumber is evident for $\sigma = 2.6$ case as well, which is less significant than the hump observed in the case of $\sigma = 1.7$  due to the spatial oscillations decaying earlier in $\sigma = 2.6$ case compared to $\sigma = 1.7$. The existence of this component in both cases further supports the idea that it stems from the characteristics of the tip vortex flow rather than from the cavity's characteristics.

The results presented in this section indicate that the spatial volume variations occurring within the growth region in the cases of $\sigma = 1.7$ and $\sigma = 2.6$ are caused by streamwise variations of the tip vortex flow field in the roll-up region, which occur within the inner regions of the tip vortex and are absent in the outer regions. Therefore, further analysis of the mean flow field is required to understand the cause of this aspect of spatial volume variations; however, this investigation falls outside the scope of the present study.



\subsubsection{Spatially-averaged Effective Radius}
\label{section:SpatialAvgReff}

\begin{figure}
\centering
\begin{subfigure}[t]{.6\textwidth}
    \centering
    \includegraphics[width=\textwidth]{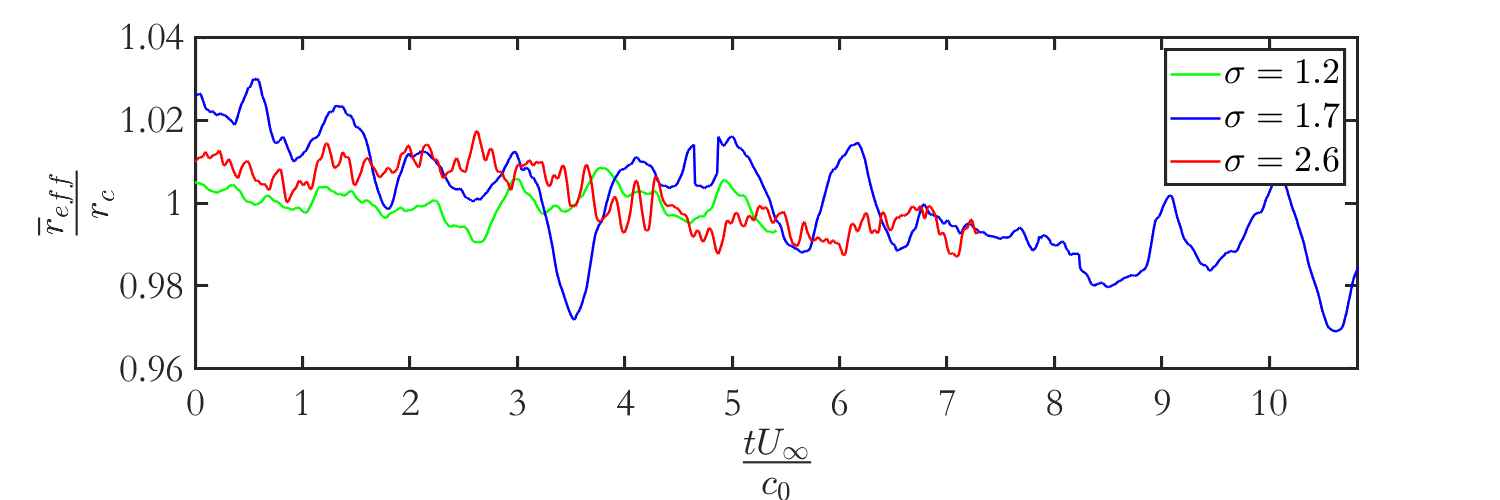}
    \caption{Non-dimensional spatially-averaged effective radius temporal development}
    \label{fig:SpaceAvgReff}
\end{subfigure}%
~
\begin{subfigure}[t]{.36\textwidth} 
    \centering
    \includegraphics[width=\textwidth]{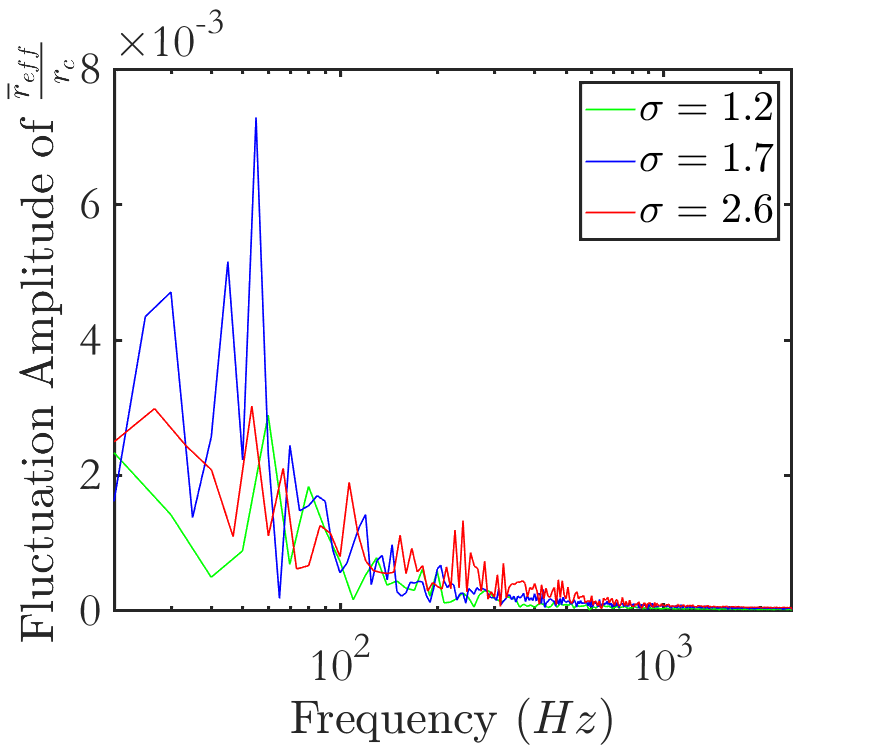}
    \caption{Frequency spectrum of the non-dimensional spatially-averaged effective radius variations}
    \label{fig:SpaceAvgReffFFT}
\end{subfigure}
\caption{Temporal dynamics of spatially-averaged effective radius}
\end{figure}

The volume of the tip vortex cavity is an important quantity that can reveal any cyclic process that the cavity might undergo. In this section, the spatially-averaged effective radius, shown as $\overline{r}_{eff}$, is employed in order to analyze the volume variations of the entire vapor cavity in the region of interest ($0.1 < x/c_0 < 1.2$). The spatially-averaged effective radius non-dimensionalized using $r_{c}$ is depicted in figure \ref{fig:SpaceAvgReff} as a function of the non-dimensional time ($t^* = \frac{t U_\infty}{c_0}$) for the three cavitation numbers. The spatial averaging for the case of $\sigma = 2.6$ is carried out for the streamwise location range of $0.1 < x/c_0 < 1.0$ because the cavity collapses between $x/c_0 = 1.0$ and $x/c_0 = 1.2$ in this case. It is evident in figure \ref{fig:SpaceAvgReff} that the volume of the cavity in the region of interest undergoes temporal variations, which occur more intensely when $\sigma = 1.7$, where cavitation occurs moderately, compared to the other cases.

The frequency spectra of the spatially-averaged effective radius fluctuations are illustrated in figure \ref{fig:SpaceAvgReffFFT} obtained using FFT algorithm. The spectra of the fluctuations demonstrate that a large proportion of the vapor volume variations occurs in the low-frequency region below $100 \, \mathrm{Hz}$ in all of the cases. Furthermore, a distinct peak is observed in the spectrum of the cavity volume variations in the case of $\sigma = 1.7$ at a frequency of $55 \, \mathrm{Hz}$, which is nearly equal to the characteristic frequency of the flow, i.e., $f_{ref} = \frac{U_\infty}{c_0} = 54.14 \, \mathrm{Hz}$. This peak in the spectrum is evident in other cases as well, which further supports the argument that this fluctuation component originates from the flow characteristic frequency. Moreover, the fluctuations' spectrum of the scenario with $\sigma = 1.7$ exhibits relatively high amplitudes at $f = 25 \mathrm{Hz}$ and $f = 30 \mathrm{Hz}$, which reveals that the spectrum has a peak between these frequencies. The  case of $\sigma = 2.6$ exhibits such a peak ($f=26.7 \, \mathrm{Hz}$) as well. Therefore, this component can be attributed to the subharmonic of the flow characteristic frequency, i.e., $\frac{f_{ref}}{2} = 27.07 \, \mathrm{Hz}$ since it is observed in two cases with different cavitation numbers.

The spectra of the spatially-averaged effective radius fluctuations of these cases differ in some other features. A distinct component at $f=45 \, \mathrm{Hz}$ is observed in the case of $\sigma = 1.7$. Moreover, the spectrum of the spatially-averaged effective radius fluctuations in the case of $\sigma = 2.6$ exhibits a local maximum at the second harmonic of the flow characteristic frequency ($\sim 107 \, \mathrm{Hz}$), which is not observed in other cases. Furthermore, in this case, the spatially-averaged effective radius possesses fluctuating components at higher frequencies ($220-240 \, \mathrm{Hz}$) as well, which is not the case in lower cavitation numbers. Since these higher frequency fluctuations are not observed in other cases, these components might be due to the specific characteristics of the cavity forming at this cavitation number. Overall, it is evident that the fluctuations of the spatially-averaged effective radius in frequencies greater than $100 \, \mathrm{Hz}$ exhibit higher intensity when $\sigma = 2.6$ compared to other cavitation numbers. Overall, it is observed that increasing the cavitation number leads to more intense vapor volume temporal variations in higher frequencies.
The results presented in this section show that the variations of vapor volume of the tip vortex cavity are dominated by the flow characteristics, and the characteristics of the cavity forming in different cases have less significant impact on overall vapor volume variations.




\subsubsection{Spatial-Temporal Characteristics of Breathing Mode Oscillations}
\label{section:Reff2DFFT}

\begin{figure}
\centering
    \begin{subfigure}[b]{\textwidth}
        \includegraphics[width=\textwidth]{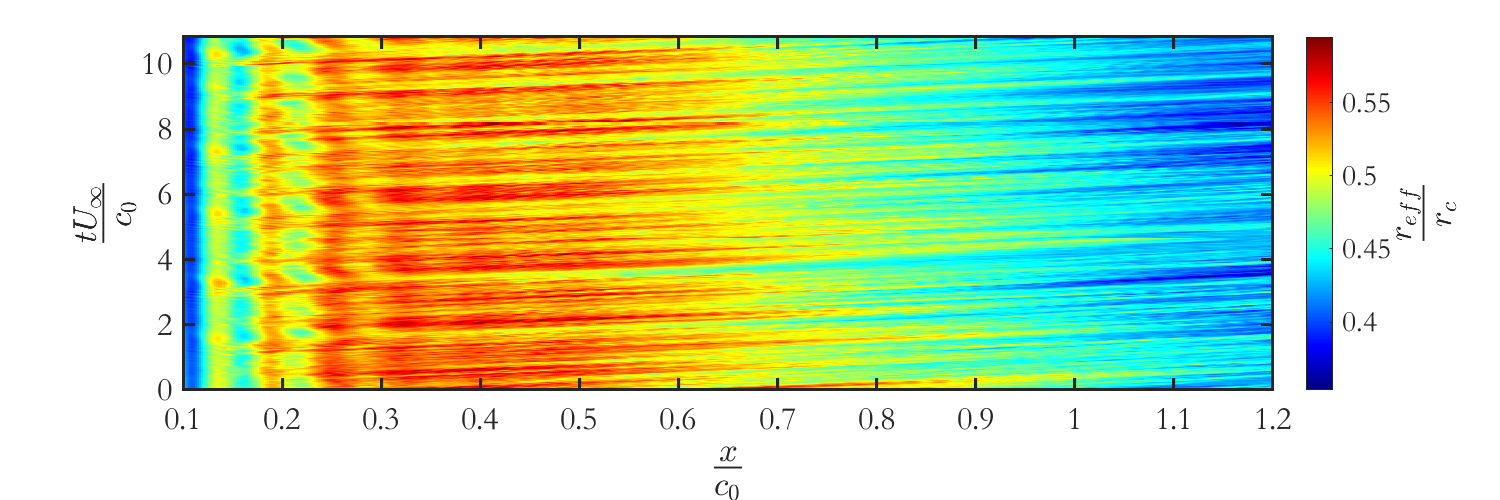}
        \caption{$\frac{r_{eff}}{r_{c}}$}
        \label{fig:ReffSpaceTime}
    \end{subfigure}
    \hfill
    \begin{subfigure}[b]{\textwidth}
        \includegraphics[width=\textwidth]{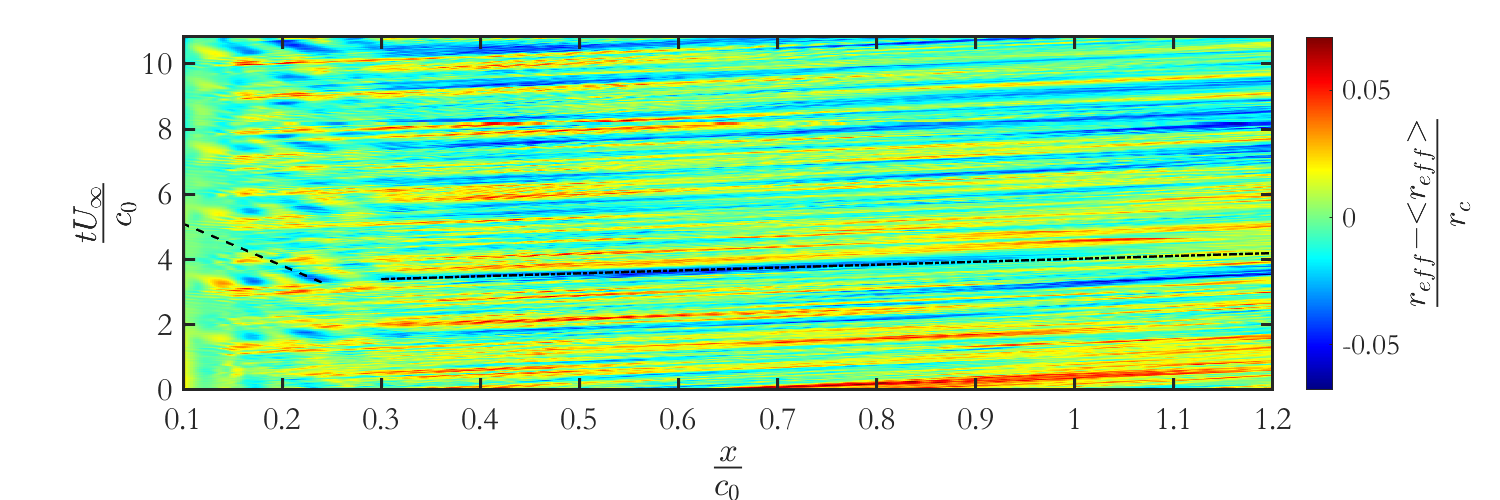}
        \centering
        \caption{$\frac{r_{eff} - <r_{eff}>}{r_{c}}$}
        \label{fig:ReffFluctSpaceTime}
    \end{subfigure}
    \caption{Spatial-temporal variations of the non-dimensional effective radius and its fluctuations ($\sigma = 1.7$)}
\end{figure}

The effective radius calculated using the method explained in \S \ref{section:ReffExtraction} is a function of time and streamwise location; therefore, it can reveal the spatial and temporal evolution of the breathing mode oscillations. The spatial and temporal variations of the non-dimensionalized effective radius for the cavitation number $\sigma = 1.7$ are illustrated in figure \ref{fig:ReffSpaceTime}, and figure \ref{fig:ReffFluctSpaceTime} displays the non-dimensional fluctuations of the effective radius at every streamwise location by subtracting the time-averaged effective radius ($<r_{eff}>$) at the corresponding location from the effective radius value. The spatial oscillations observed in this case in the cavity growth region $0.1 < x/c_0 < 0.4$ can be observed in figure \ref{fig:ReffSpaceTime}. 

Furthermore, it is evident in figure \ref{fig:ReffFluctSpaceTime} that the behavior of the effective radius fluctuations alters at $x/c_0 \sim 0.3$ in this case. It is apparent in this figure that in the upstream region of $x/c_0 = 0.3$, some of the peaks and troughs of effective radius fluctuations visible as red and blue stripes, respectively, propagate upstream (shown as a dashed line in figure \ref{fig:ReffFluctSpaceTime}), in contrast to downstream of $x/c_0 =0.3$ where such disturbances propagate downstream (shown as a dot-dashed line in figure \ref{fig:ReffFluctSpaceTime}). The slope of these upstream and downstream propagating disturbances reveals that the propagation speed of the disturbances propagating downstream is greater than that of the ones propagating upstream in the region $x/c_0 < 0.3$. These differences reveal the change of cavity oscillatory behavior as it advances from the growth region to the decay region. Moreover, it can be observed in figure \ref{fig:ReffSpaceTime} that at a streamwise location of $x/c_0 \sim 0.22$, the effective radius peaks periodically approximately every $t_{ref} = \frac{c_0}{U_\infty}$ (flow characteristic time scale), which indicates a fluctuating component of the effective radius at a frequency of $f_{ref} = \frac{U_\infty}{c_0}$ at this location.

\begin{figure}
\centering
\begin{minipage}[t]{.48\textwidth}
    \centering
    \includegraphics[width=\textwidth]{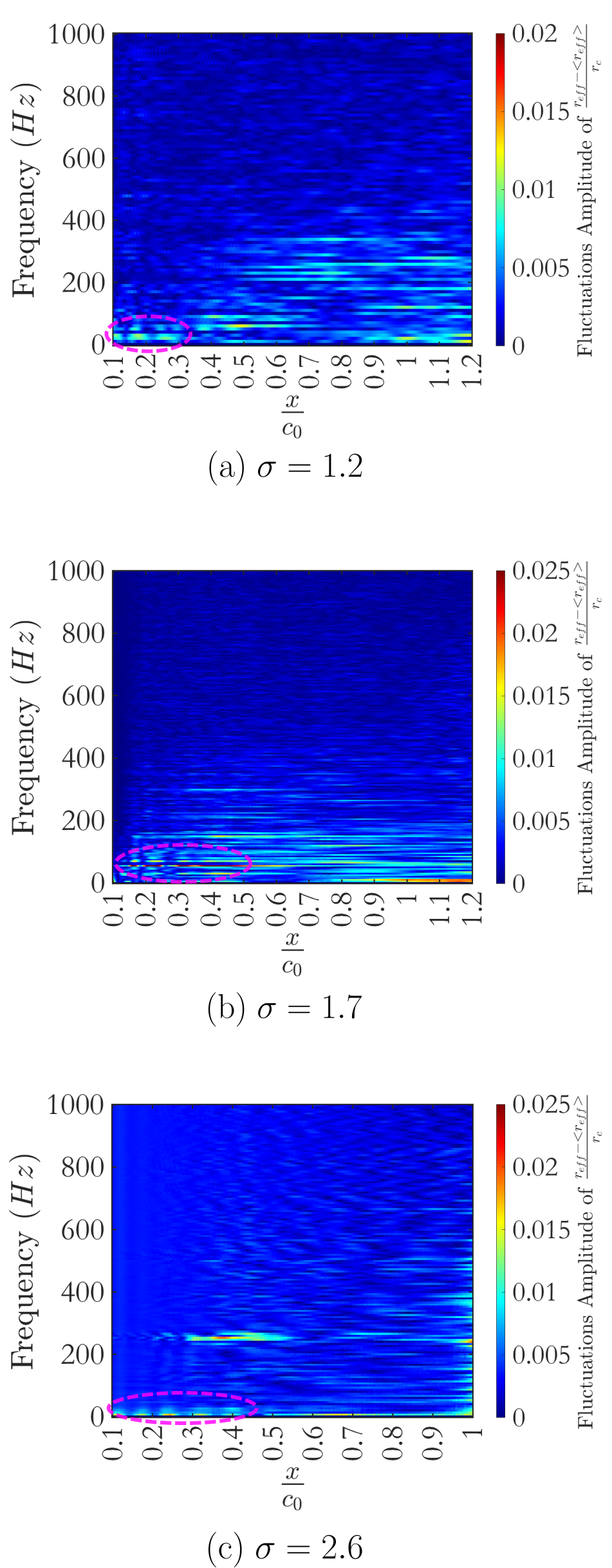}
    \caption{Spectrum of the effective radius fluctuations with respect to downstream location}
    \label{fig:ReffFluctSpectrum}
\end{minipage}%
\hfill
\begin{minipage}[t]{.48\textwidth} 
    \centering
    \includegraphics[width=\textwidth]{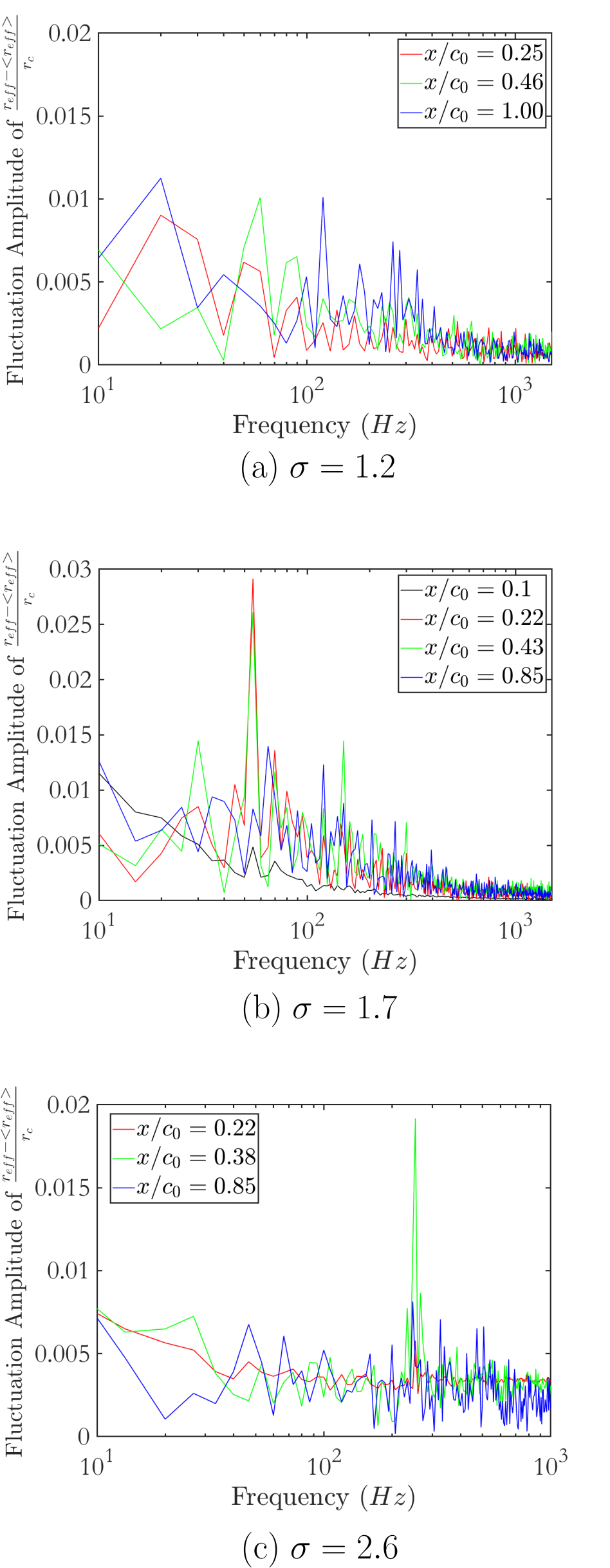}
    \caption{Frequency spectrum of the effective radius oscillations at two streamwise locations}
    \label{fig:ReffFFTX}
\end{minipage}
\end{figure}

The spectra of the effective radius fluctuations as a function of streamwise location at different cavitation numbers are shown in figure \ref{fig:ReffFluctSpectrum}. It is evident in this figure that similar to the behavior of the spatially-averaged effective radius fluctuations presented in \S \ref{section:SpatialAvgReff}, the major proportion of the local fluctuations of the effective radius occurs at low frequencies (below $300 \, \mathrm{Hz}$). Furthermore, it is evident in this figure that breathing mode oscillation characteristics alter as the cavity travels away from the tip. Another characteristic of the breathing mode oscillations observed in the spectra of all of the cases is the regions of high-amplitude fluctuations within the growth region which vanish and appear periodically with streamwise location. These regions are specified in figure \ref{fig:ReffFluctSpectrum} using magenta dashed ellipses. Overall, according to this figure, the breathing mode oscillations occur at lower frequencies in the growth region and higher frequency components of breathing mode oscillation intensify as the cavity travels downstream. 

The frequency spectra of breathing mode oscillations at three streamwise locations at different cavitation numbers are illustrated in figure \ref{fig:ReffFFTX}. In the case of $\sigma = 1.7$, the breathing mode oscillations spectrum at $x/c_0 = 0.1$ is also plotted, which reveals that the breathing mode oscillations are small in amplitude at this location, and the intensification of the breathing mode oscillations is apparent when the spectrum at this location is compared to locations further downstream. This indicates that these breathing mode oscillations are not initially present as the cavity forms and are introduced onto the cavity surface as it flows downstream. Moreover, a distinct peak is observed at $f=55 \, \mathrm{Hz}$ at both $x/c_0 = 0.22$ and $x/c_0 = 0.43$. The spectrum at $x/c_0 = 0.43$ exhibits two other local maxima at $f=150 \, \mathrm{Hz}$ and $300 \, \mathrm{Hz}$. As the cavity travels further downstream and develops, its spectrum displays more peaks, an example of which is the spectrum at $x/c_0 = 0.85$, with the most significant maximum amplitudes occurring at $f=65\,\mathrm{Hz}$, $120\,\mathrm{Hz}$, $150\,\mathrm{Hz}$, $205\,\mathrm{Hz}$, and $265\,\mathrm{Hz}$. This further supports that fluctuating components at higher frequencies intensify as the cavity advances from the growth region to the decay region.

In the case of $\sigma = 1.2$, no distinct significant peak is evident in the spectrum. Comparison of the spectrum at $x/c_0 = 1.00$ with further upstream locations demonstrates the intensification of breathing mode oscillations as the cavity progresses from the growth region to the decay region in this case as well. The peaks observed in this case at $x/c_0 = 1.00$ possess frequencies of $f = 20 \, \mathrm{Hz}$, $120 \, \mathrm{Hz}$, $260 \, \mathrm{Hz}$, $280 \, \mathrm{Hz}$, $340 \, \mathrm{Hz}$.

%
At a cavitation number of $\sigma = 2.6$, the oscillatory behavior alters from that of lower cavitation numbers. The low-frequency oscillations are less intense, and there is a distinct peak at $f = 253 \, \mathrm{Hz}$ in the spectrum within the transition region. Overall, it can be observed that in this case, the breathing mode spectrum extends to higher frequencies compared to the cases with lower cavitation numbers.

\begin{figure}
    \centering
    \includegraphics[width=\textwidth]{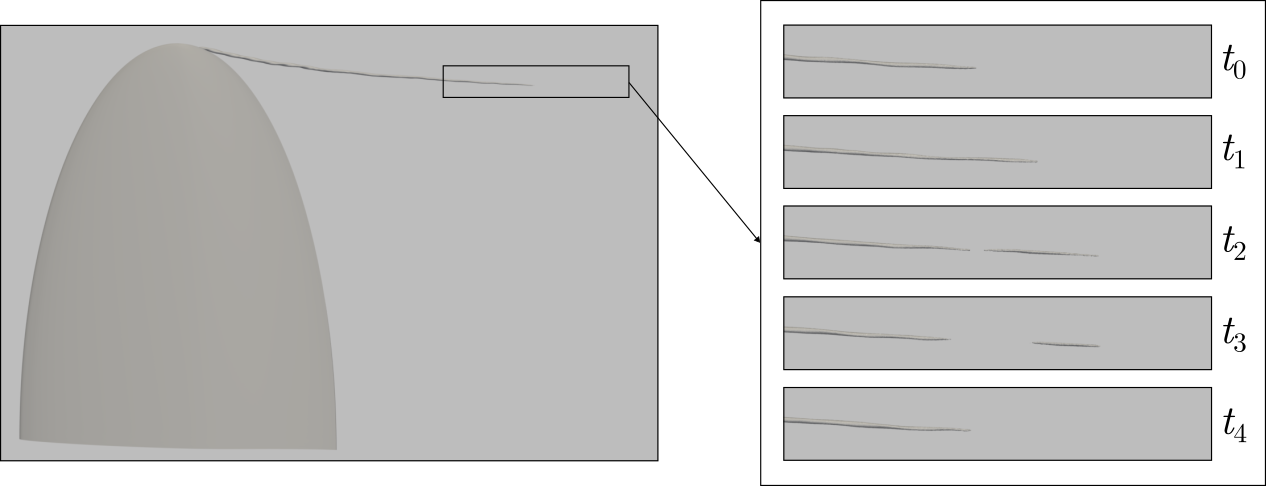}
    \caption{The cyclic behavior of tip vortex cavity's front ($\sigma = 2.6$)}
    \label{fig:TVCCollapse}
\end{figure}

Since the fine mesh region encompasses the entire cavity in the case of $\sigma = 2.6$, the end effects of the tip vortex cavity can be examined in this case. The results indicate that the cavity undergoes a cyclic pattern of growth and detachment at its front, which is illustrated in figure \ref{fig:TVCCollapse}. The cycle starts with elongation of the cavity, which progresses until it reaches a critical length, followed by detachment of a part of the cavity and shedding downstream. After the detachment, the tip vortex cavity's end retracts until it reaches a minimum length, and the cycle starts over with elongation. The frequency of this cycle was found to be within the range of $250 \, \mathrm{Hz} < f < 300 \, \mathrm{Hz}$, which is similar to the peak frequencies observed in the breathing mode oscillation spectrum in figure \ref{fig:ReffFluctSpectrum}c between $x/c_0 = 0.3$ and $x/c_0 = 0.6$. This highlights a potential correlation between the breathing mode oscillation within this frequency range and the cyclic pattern of growth and detachment observed in the tip vortex cavity's front.

\subsubsection{Decomposition of Breathing Mode Oscillations}
\label{section:ReffPOD}
The analysis of the effective radius reveals that the breathing mode oscillations exhibit various characteristics in different regions of the cavity. In this section, Proper Orthogonal Decomposition (POD) is employed to decompose the effective radius data and obtain the main breathing mode oscillations. The time-averaged effective radius is subtracted from the results to extract the fluctuating components. The analysis in this section is only carried out for the case of $\sigma = 1.7$.

\begin{figure}
\centering
    \begin{subfigure}[b]{0.78\textwidth}
        \includegraphics[width=\textwidth]{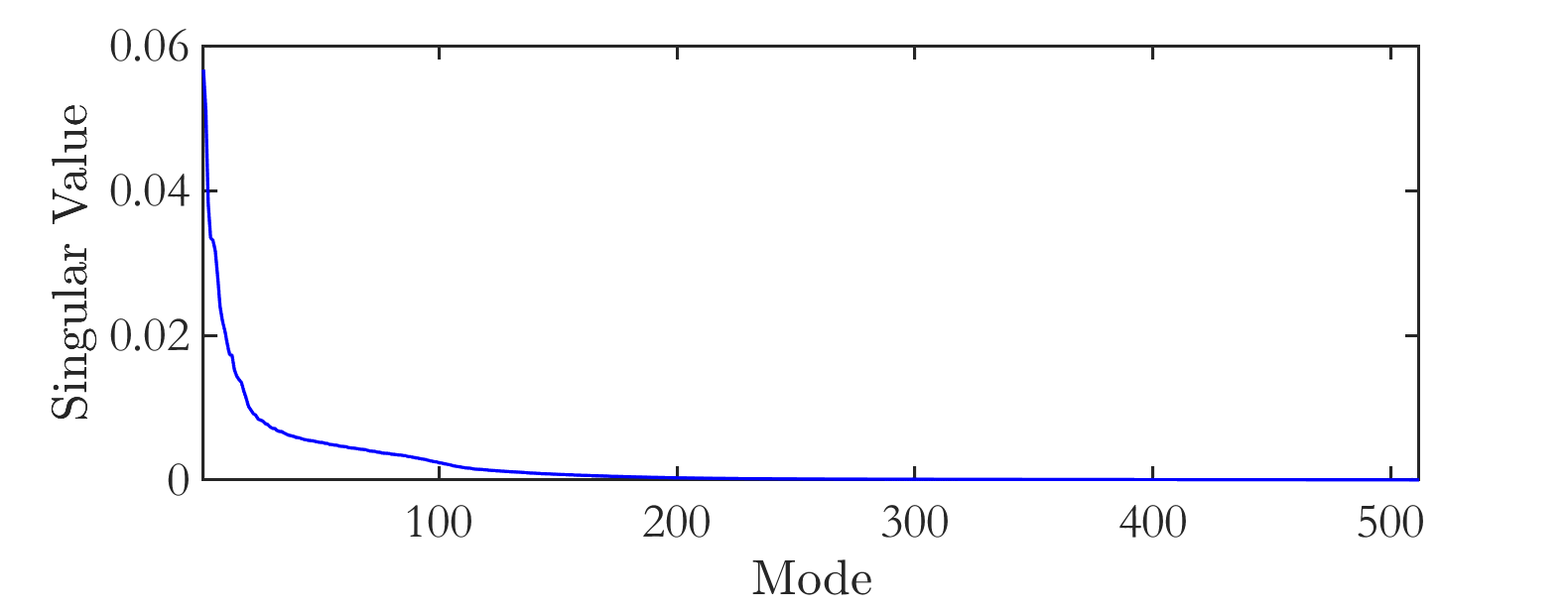}
        \caption{Singular values of the effective radius fluctuation modes}
        \label{fig:ReffSingularValues}
    \end{subfigure}
    \hfill
    \begin{subfigure}[b]{0.78\textwidth}
        \includegraphics[width=\textwidth]{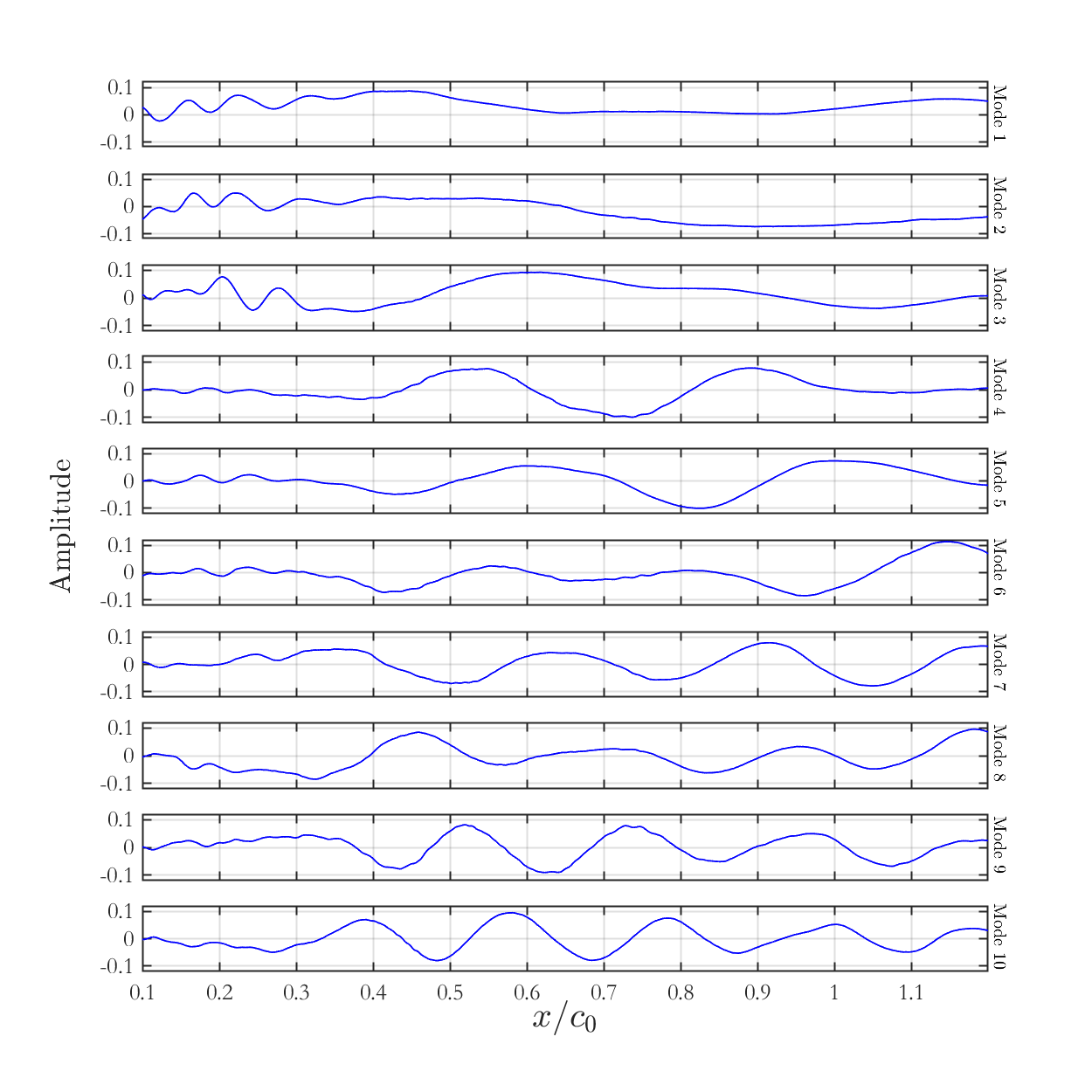}
        \centering
        \caption{First 10 modes of effective radius fluctuations obtained using POD}
        \label{fig:ReffPODModes}
    \end{subfigure}
    \caption{Proper orthogonal decomposition of breathing mode oscillations ($\sigma = 1.7$)}
\end{figure}

The energy contribution of the effective radius fluctuation modes is plotted in figure \ref{fig:ReffSingularValues}, which shows that the fluctuations are distributed over a large number of modes; however since the goal here is to gain insight into the fluctuations and not reduced-order modeling, the first few modes can be employed. Figure \ref{fig:ReffPODModes} depicts the first 10 modes of the effective radius fluctuations. The change of behavior downstream of the cavity growth region, which corresponds to $x > 0.4$, is apparent in the fluctuation modes. It is also evident from the modes that the fluctuations in the cavity growth region have relatively small wavelengths and amplitudes compared to the decay region, which is in line with the lower propagation speed observed within the growth region compared to the decay region as shown in figure \ref{fig:ReffFluctSpectrum}.

The amplification and the alteration in the behavior of the effective radius fluctuations downstream of the cavity growth region indicate that some oscillations are introduced onto the cavity interface in the roll-up process as mentioned before. The source of these perturbations should be sought in the temporal variations of the tip vortex flow field and the roll-up of the wake into the tip vortex flow, which is not within the scope of this study.


\subsection{Effect of Cavity Surface Oscillations on Pressure Fluctuations}
\label{section:PressureFluctuations}
\begin{figure}
\centering
\begin{minipage}[t]{.33\textwidth}
    \centering
    \includegraphics[width=\textwidth]{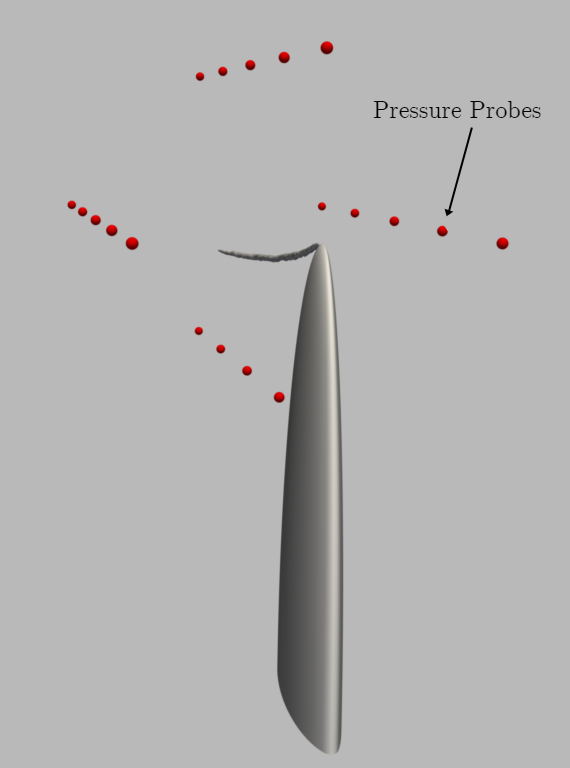}
    \caption{Locations of pressure probing}
    \label{fig:PressureProbes}
\end{minipage}%
\hfill
\begin{minipage}[t]{.62\textwidth} 
    \centering
    \includegraphics[width=\textwidth]{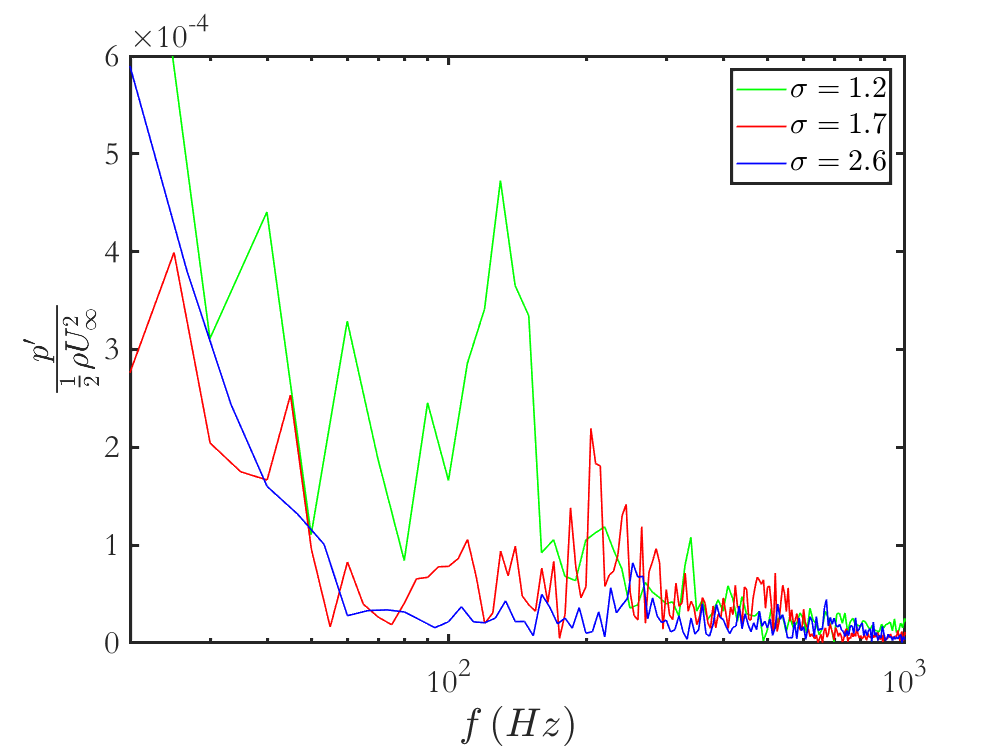}
    \caption{Spectrum of the pressure fluctuations at point ($x=0, \, y=0, \, z=h+0.5c_0$)}
    \label{fig:FFTPressure}
\end{minipage}
\end{figure}

In order to investigate the contribution of cavity surface oscillations to the pressure fluctuations in different cases, pressure is probed at 19 points within the domain, as shown in figure \ref{fig:PressureProbes}, on which FFT is performed subsequently to obtain the spectra of the pressure fluctuations. The spectra of the pressure fluctuations at these points exhibit similar behavior in all of the cases; therefore, the pressure fluctuations spectrum at only one point, which is located $0.5c_0$ away from the tip in the spanwise direction ($x=0, \, y=0, \, z=h+0.5c_0$), is reported herein and is shown in figure \ref{fig:FFTPressure} for the three cases.



It is evident in figure \ref{fig:FFTPressure} that decreasing the cavitation number leads to more intense pressure fluctuations. This indicates that tip vortex cavitation is, indeed, contributing to the pressure fluctuations within the domain. Moreover, it can be observed that as the cavitation number decreases, the spectrum shifts towards lower frequencies overall. The pressure fluctuations at low frequencies (below $50 \mathrm{Hz}$) are relatively significant in all cases. Besides these general features, the pressure fluctuations spectrum of each case exhibits some unique characteristics as well.

The pressure fluctuations spectrum in the case of $\sigma = 1.2$ displays a high-amplitude peak at a frequency of $f = 130 \, \mathrm{Hz}$, in addition to other local maxima at $f = 40 \, \mathrm{Hz}$, $60 \, \mathrm{Hz}$, $220 \, \mathrm{Hz}$, and $340 \, \mathrm{Hz}$. It is evident in figures \ref{fig:ReffFluctSpectrum}a and \ref{fig:ReffFFTX}a that a high-amplitude component is present in the breathing mode spectrum at $f = 60 \, \mathrm{Hz}$ within the growth region. This peak is also observed in the spatially-averaged effective radius spectrum in this case illustrated in figure \ref{fig:SpaceAvgReffFFT}. The maximum observed in the pressure fluctuations spectrum at $f = 130 \, \mathrm{Hz}$ is the peak of a hump in the frequency range of $80-160 \, \mathrm{Hz}$. The frequency of the center of this hump, i.e., $120 \, \mathrm{Hz}$, is observed in figure \ref{fig:ReffFluctSpectrum}a which shows relatively large breathing mode oscillations at this frequency downstream of $x/c_0 = 0.85$. Moreover, a peak is observed in the spatially-averaged effective radius spectrum depicted in figure \ref{fig:SpaceAvgReffFFT} at $f = 130 \, \mathrm{Hz}$. Figure \ref{fig:ReffFluctSpectrum}a also indicates the presence of breathing mode oscillations at a frequency of $340 \, \mathrm{Hz}$ within the developed region of the cavity in the case of $\sigma = 1.2$.

The pressure fluctuations spectrum in the case with a cavitation number of $\sigma = 1.7$ indicates the presence of low-frequency peaks at $f \sim 25 \, \mathrm{Hz}$ and $f \sim 45 \, \mathrm{Hz}$.  The spectrum exhibits a relatively broadband hump in the frequency range $160-280 \, \mathrm{Hz}$, with a peak at $f \sim 205 \, \mathrm{Hz}$ and less significant peaks at $185 \, \mathrm{Hz}$, $245 \, \mathrm{Hz}$, $265 \, \mathrm{Hz}$, and $285 \, \mathrm{Hz}$. These peaks are present in the pressure fluctuations spectra at all of the pressure probing locations, which indicates that these pressure fluctuations are generated by a monopole-type source. The low-frequency peaks at $f \sim 25 \, \mathrm{Hz}$ and $f \sim 45 \, \mathrm{Hz}$ were also observed in the spatially-averaged effective radius spectrum depicted in figure \ref{fig:SpaceAvgReffFFT}; however, the cavity volume variations at $f \sim 55\,\mathrm{Hz}$ seem to not be correlated with the pressure fluctuations. Moreover, a local maximum at $f = 205 \, \mathrm{Hz}$ is also evident in the spatially-averaged effective radius spectrum in figure \ref{fig:SpaceAvgReffFFT}.


Comparing the effective radius fluctuations spectra presented in figure \ref{fig:ReffFFTX}b with the pressure fluctuations spectrum indicates the presence of the distinct fluctuating components at $f=205 \, \mathrm{Hz}$ and $f=265 \, \mathrm{Hz}$ in the effective radius fluctuations spectrum of the location within the developed region ($x/c_0 = 0.85$); however, this peak is not observed in the breathing mode spectrum at $x/c_0 = 0.22$, which is in the cavity formation region. On the contrary, the peak at $f=45 \, \mathrm{Hz}$ in the pressure fluctuations spectrum is also observed in the breathing mode spectrum at $x/c_0 = 0.22$ and not in the spectrum at $x/c_0 = 0.85$.

In the case of $\sigma = 2.6$, the pressure fluctuations spectrum exhibits a relatively small hump within the frequency range of $220-300 \, \mathrm{Hz}$ with a peak at $f = 253 \, \mathrm{Hz}$. The breathing mode oscillations spectrum for this case depicted in figure \ref{fig:ReffFluctSpectrum}c revealed intense breathing mode oscillations at $f = 253 \, \mathrm{Hz}$ within the streamwise location range of $0.28 \, x/c_0 \, 0.55$. Moreover, the spatially-averaged effective radius spectrum presented in figure \ref{fig:SpaceAvgReffFFT} exhibits some local maxima at $f = 227 \, \mathrm{Hz}$, $240 \, \mathrm{Hz}$, $253 \, \mathrm{Hz}$, and $280 \, \mathrm{Hz}$. Furthermore, the cyclic growth and detachment behavior of the cavity's end was found to be occurring within a frequency range of $250-300 \, \mathrm{Hz}$, which indicates that this cyclic tail effect is potentially influencing the pressure fluctuations within the domain as well. Figure \ref{fig:ReffFluctSpectrum}c also indicates the presence of local breathing mode oscillations within the growth region and upstream of the decay region in this case at low frequencies from $10-40 \, \mathrm{Hz}$. These oscillations might be correlated with higher amplitudes of pressure fluctuations at frequencies lower than $40 \, \mathrm{Hz}$ in this case compared to the case of $\sigma = 1.7$, wherein such local oscillations within the growth region occur at higher frequencies.

Based on the correlations between the cavity breathing mode oscillations and the pressure fluctuations, it can be concluded that the growth region of the cavity is correlated with the low-frequency peaks observed in the pressure fluctuations spectrum, and the mid-range of this spectrum, where the hump in the pressure fluctuations spectrum occurs as well, is correlated with the breathing mode oscillations in the decay region of the tip vortex cavity.

\section{Concluding remarks}
\label{section:conclusion}

In this study, tip vortex cavitating flow over a stationary elliptical NACA66(2)-415 hydrofoil was numerically investigated using a finite element-based cavitation flow solver. To address the lack of general mesh resolution requirements for TVC simulation, a pressure gradient-based length scale was developed employing the Rankine vortex model, Kutta-Jukowski theorem, and the turbulent boundary layer thickness. This new length scale was then utilized for non-dimensionalization of the mesh resolution within the tip vortex flow region. The simulations carried out in this study for various non-dimensional mesh resolution values ($\Delta r ^ *$) revealed that a $\Delta r ^ *$ of 18.5 or lower should be utilized for the accurate simulation of tip vortex cavitating flows using LES.

The oscillatory dynamics of the tip vortex cavity were investigated in the results obtained from the simulations carried out with a mesh satisfying the proposed resolution requirements. The simulations were conducted for three different cases which differed only in the cavitation number. The cavity dynamics captured in the simulations in this study were shown to agree well with the analytical solution.

In the next step, since it is well-known that the breathing mode of oscillation is the most influential mode of tip vortex cavity surface oscillations concerning the generated noise compared to other modes, a new method was developed for the extraction of this mode of oscillation from the numerical results using a new parameter named the cavity effective radius ($r_{eff}$). Investigation of the temporally-averaged effective radius demonstrated that the tip vortex cavity experiences different regions as it progresses away from the tip. The results indicated that the growth region, i.e., the region close to the tip where the cavity is forming and experiences an overall increasing trend in volume, shrinks with increasing the cavitation number. Furthermore, it was demonstrated that the decay occurs at an increased slope in scenarios with higher cavitation numbers. Another observation was the occurrence of spatial periodic cavity volume variations within the growth region. These variations exhibit similar dominant wavenumbers between cases with different cavitation numbers, which reveals that the dominant wavenumber of such oscillations is related to the mean tip vortex roll-up process and not the characteristics of the cavity forming in each specific case.

Investigation of the spatially-averaged effective radius indicated that the overall vapor volume undergoes temporal variations, with the most intense oscillations occurring in the moderately cavitating case, i.e., $\sigma = 1.7$. The analysis revealed that the flow characteristic frequency and its subharmonic dominate the overall vapor volume variations, with the most significant peak having a frequency similar to the flow characteristic frequency in all of the cases. Moreover, the results revealed that increasing the cavitation numbers leads to more intense vapor volume variations at higher frequencies, with the most intense high-frequency vapor volume variations occurring when $\sigma = 2.6$.

Subsequently, the local breathing mode oscillations were investigated. The local breathing mode oscillations exhibit different propagation characteristics in the growth and decay regions in terms of speed and direction. More evidence was provided for the alteration of the cavity's oscillatory behavior between the growth and decay regions employing Proper Orthogonal Decomposition. Moreover, the results indicated that the breathing mode oscillations are not inherent to the tip vortex cavity, and such oscillations appear and intensify as the cavity travels away from the tip. Therefore, the tip vortex cavity displays different characteristics in the growth and decay regions. The breathing mode oscillations within the decay region were observed to extend over a more broadband range of frequencies compared to the growth region. Overall, the breathing mode oscillations within the growth region were observed to occur within the low-frequency range, and such oscillations within the decay region were found to possess higher frequencies. The simulations revealed that at a low cavitation number, similar to the case of $\sigma = 1.2$ in this work, no significant peak was observed in the breathing mode spectrum; however, distinct peaks were observed at higher cavitation numbers with the most intense ones occurring in the case with $\sigma = 1.7$. Furthermore, similar to the spatially-averaged effective radius fluctuation spectrum, the local breathing mode oscillation in the case of $\sigma = 2.6$ extended to higher frequencies compared to lower cavitation numbers.

An interesting feature of the tip vortex cavity was observed when $\sigma = 2.6$, which was the cyclic growth and detachment behavior of the cavity at its end. It was observed that the cavity undergoes a process of elongation, detachment, and retraction at its end, which was found to occur within a frequency range of $250-300 \, \mathrm{Hz}$. The local breathing mode oscillations in this working condition also displayed fluctuations within this frequency range, which indicates a potential correlation between the cyclic tail behavior of the tip vortex cavity and its breathing mode oscillations. This aspect of tip vortex cavitation requires further investigation.

The pressure fluctuations within the domain were investigated in cases with different cavitation numbers. The results indicated that decreasing the cavitation number, which leads to a higher extent of tip vortex cavitation, results in higher levels of pressure fluctuations overall. Another effect of decreasing the cavitation number was found to be shifting the pressure fluctuations spectrum towards lower frequencies, which is similar to the behavior of breathing mode oscillations. Comparison of the pressure fluctuations spectra at different cavitation numbers with the overall vapor volume variations and local breathing mode oscillations at the corresponding cavitation numbers demonstrated the existence of correlations between the pressure fluctuations and the breathing mode of oscillation. The center frequencies of the humps observed in the pressure fluctuations spectra were found to be related to the breathing mode oscillations spectrum in all of the cases.

Despite the existence of these correlations, there were some features in the pressure fluctuations that were not related to the breathing mode of cavity surface oscillation. Moreover, some significant fluctuating components were observed in the breathing mode oscillations which did not affect the pressure fluctuations despite having high amplitudes. Therefore, further investigation is required to thoroughly understand the correlations between the breathing mode oscillations and the pressure fluctuations. 

\bibliographystyle{jfm}
\bibliography{main}

\end{document}